\documentclass[10pt,journal,compsoc]{IEEEtran}
\usepackage{amsmath,amsfonts}
\usepackage{algorithmic}
\usepackage{array}
\usepackage[caption=false,font=normalsize,labelfont=sf,textfont=sf]{subfig}
\usepackage{textcomp}
\usepackage{stfloats}
\usepackage{url}
\usepackage{verbatim}
\usepackage{graphicx}
\def\BibTeX{{\rm B\kern-.05em{\sc i\kern-.025em b}\kern-.08em
    T\kern-.1667em\lower.7ex\hbox{E}\kern-.125emX}}
\usepackage{balance}

\usepackage{ragged2e}

\usepackage{hyperref}
\usepackage{graphicx}
\usepackage{orcidlink}

\usepackage{booktabs}
\usepackage{amsmath,amsfonts}
\usepackage{algorithmic}
\usepackage{graphicx}
\usepackage{textcomp}
\usepackage{xcolor}
\usepackage[normalem]{ulem}
\usepackage{bbding}
\usepackage{url}
\usepackage{graphicx}
\usepackage{xcolor}
\usepackage{colorspace}

\begin{document}
\title{You Can Wash Hands Better: Accurate Daily Handwashing Assessment with a Smartwatch}
% \author{IEEE Publication Technology Department
% \thanks{Manuscript created October, 2020; This work was developed by the IEEE Publication Technology Department. This work is distributed under the \LaTeX \ Project Public License (LPPL) ( http://www.latex-project.org/ ) version 1.3. A copy of the LPPL, version 1.3, is included in the base \LaTeX \ documentation of all distributions of \LaTeX \ released 2003/12/01 or later. The opinions expressed here are entirely that of the author. No warranty is expressed or implied. User assumes all risk.}}

\author{Fei Wang\orcidlink{0000-0002-0750-6990}, Tingting Zhang\orcidlink{0009-0007-2157-4360}, Xilei Wu\orcidlink{0000-0001-6482-2693},  Pengcheng Wang\orcidlink{0009-0004-5397-9811} \\ Xin Wang\orcidlink{0000-0003-4480-1910}, Han Ding\orcidlink{0000-0002-5274-7988}, Jingang Shi\orcidlink{0000-0001-7070-6365}, Jinsong Han\orcidlink{0000-0001-5064-1955}, Dong Huang\orcidlink{0000-0001-9798-0081}
\thanks{This work was supported by the National Natural Science Foundation of China under grant 62102307, U21A20462, 62372400, 62372365, 62311530046, 62002283, and Fundamental Research Funds for the Central Universities.\\Fei Wang, Tingting Zhang, Xin Wang, Pengcheng Wang, and Jingang Shi are with the School of Software Engineering, Xi'an Jiaotong University, Xi'an 710049, China. Fei Wang is also with the State Key Laboratory of Integrated Services Networks, Xidian University, Xi'an 710071, China. Xilei Wu is with the Local Life Research and Development Center, Kuaishou Technology Inc., Beijing 100085, China. 
Han Ding is with the School of Computer Science and Technology, Xi'an Jiaotong University, Xi'an 710049, China. Jinsong Han is with the College of Computer Science and Technology, Zhejiang University, Hangzhou 310027, China. Dong Huang is with the Robotics Institute, Carnegie Mellon University, Pittsburgh 15213, USA.
\\ Fei Wang~(email: feynmanw@xjtu.edu.cn) and Jingang Shi~(email: jingang@xjtu.edu.cn) are corresponding authors.}
}

\markboth{IEEE Transactions on Mobile Computing,~Vol.~00, No.~0, May~2025}%
{You Can Wash Hands Better: Accurate Daily Handwashing Assessment with a Smartwatch}

\IEEEtitleabstractindextext{

\begin{abstract}
\justifying Hand hygiene is among the most effective daily practices for preventing infectious diseases such as influenza, malaria, and skin infections. While professional guidelines emphasize proper handwashing to reduce the risk of viral infections, surveys reveal that adherence to these recommendations remains low. To address this gap, we propose UWash, a wearable solution leveraging smartwatches to evaluate handwashing procedures, aiming to raise awareness and cultivate high-quality handwashing habits. We frame the task of handwashing assessment as an action segmentation problem, similar to those in computer vision, and introduce a simple yet efficient two-stream UNet-like network to achieve this goal. Experiments involving 51 subjects demonstrate that UWash achieves 92.27\% accuracy in handwashing gesture recognition, an error of $<$0.5 seconds in onset/offset detection, and an error of $<$5 points in gesture scoring under user-dependent settings. The system also performs robustly in user-independent and user-independent-location-independent evaluations. Remarkably, UWash maintains high performance in real-world tests, including evaluations with 10 random passersby at a hospital 9 months later and 10 passersby in an in-the-wild test conducted 2 years later. UWash is the first system to score handwashing quality based on gesture sequences, offering actionable guidance for improving daily hand hygiene. The code and dataset are publicly available at \url{https://github.com/aiotgroup/UWash}.

\end{abstract}

}
\maketitle

\section{Introduction}~\label{sec:introduction}

\IEEEPARstart{H}and hygiene is an efficient and effective approach to preventing various infectious diseases, e.g., colds and flu, enterovirus infections, and skin infections. We conducted a questionnaire on handwashing knowledge and practices over 505 subjects across 26 provinces in China. The questionnaire shows that 96.04\% of subjects have heard of World Health Organization~(WHO) guidelines or 7-step guidelines but only 34.65\% of subjects follow the guidelines, similar to the situation reported in two other surveys conducted in Germany~\cite{mieth2021they} and in Nigeria~\cite{wada2021safe}. Thus, it is critical to raise people's awareness and cultivate habits of high-quality handwashing in daily life. 

\begin{figure}[t]
    \centering
    \includegraphics[width=1\linewidth]{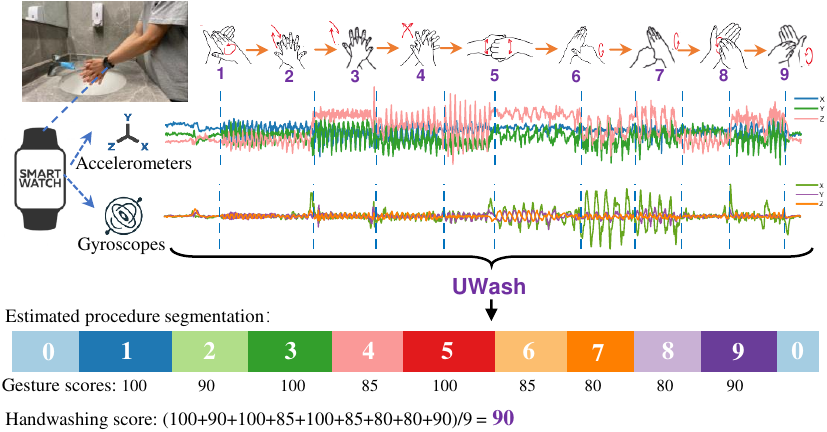}
    \caption{UWash utilizes records of motion sensors of smartwatches, i.e., accelerometers and gyroscopes, to segment handwashing gestures, estimate the duration of each gesture, and score the quality of each gesture as well as the entire procedure with WHO guidelines.}
    \label{fig:overview}
\end{figure}

In healthcare scenarios, hospitals and healthcare centers would like to employ auditors to directly observe and assess the handwashing procedures to promote the healthcare workers'~(HCWs) adherence to the guidelines. However, this approach is time-consuming, time-delayed, and costly. For these concerns, many works propose automatic handwashing monitoring systems to promote HCWs' adherence and awareness~\cite{kinsella2007electronic, pineles2014accuracy, rabeek2014reliable, edmond2010successful, llorca2011vision, zhong2016washindepth, haque2017towards, li2018wristwash, khamis2020rfwash, samyoun2021iwash}. Given the above practices, we believe the accessibility of an automatic handwashing monitoring solution in people's daily life is the next step to promoting ordinary people's awareness of handwashing. 
Recent advancements in wearable technology have spurred interest in portable handwashing assessment systems. Solutions like WristWash~\cite{li2018wristwash} and iWash~\cite{samyoun2021iwash} use wrist-worn devices to evaluate handwashing gestures but rely on predefined motion patterns or manual activation, which reduce usability. Similarly, WashRing~\cite{xu2024washring}, employing smart rings, faces issues with potential interference during handwashing. AWash~\cite{cao2021awash} and RFWash~\cite{khamis2020rfwash}, on the other hand, utilize infrastructure-based triggers, such as smart faucets or RFID sensors, but lack portability and adaptability across environments.

To address these limitations, We propose UWash, an automatic solution with smartwatches, to assess handwashing techniques for the purpose of raising people's awareness and cultivating habits of high-quality handwashing. As shown in Fig.~\ref{fig:overview}, UWash leverages the inertial measurement units~(IMUs) of the smartwatch, i.e., accelerometers and gyroscopes, to detect the start/end time of the handwashing procedures, estimate the duration of each gesture, score the quality of estimated gestures in accordance with the WHO guidelines, and further score the quality of the entire handwashing procedure. With the estimated duration and scores, users could check their handwashing procedures and promote their handwashing techniques accordingly. Compared with existing works, our UWash provides six key features in the usage aspect and functionality aspect that enable it suitable for daily use.

\begin{itemize}
    \item  \textbf{Flexible usage across various locations.} UWash is designed to assess handwashing procedures in various environments, including homes, workplaces, and restaurants, offering flexibility for daily use. In contrast, most of the existing approaches can only work at fixed places where systems are deployed. For example, alcohol sensors~\cite{edmond2010successful}, pressure sensors~\cite{kinsella2007electronic}, ultrasonic hot spots~\cite{rabeek2014reliable}, and RFID~\cite{pineles2014accuracy} are embedded into dispensers to count handwashing;  cameras~\cite{llorca2011vision, zhong2016washindepth, haque2017towards, wang2022handwashing, nagar2022hand, asif2024resmfuse} and mmWave devices~\cite{khamis2020rfwash} are placed on walls to estimate handwashing. 
    
    \item \textbf{Independent operation without additional sensors.} UWash relies solely on the IMU of smartwatches, eliminating the need for any extra sensors. In contrast, many existing systems need additional sensors, such as Bluetooth beacons, RFID cards, and Wi-Fi modules on dispensers or on the door, as triggers to detect people's approaching and start to work~\cite{mondol2015harmony,zhong2016washindepth,samyoun2021iwash,cao2021awash, shrimali2023novel}.

    \item \textbf{Automatic detection of start and end.}
    UWash can automatically identify the onset and offset of handwashing procedures. In contrast, the official handwashing APP of the Apple Watch and iWash~\cite{samyoun2021iwash} require user activation through clicks or voice commands, which can decrease user willingness and frequency of use. AWash~\cite{cao2021awash} leverages the smart faucet or smart foam soap dispenser to detect the onset and offset of handwashing procedures, limiting the assessment places.

    \item \textbf{Flexible for random gesture sequences.} Fig.~\ref{fig:overview} shows 9 gestures of WHO guidelines. Some approaches require the users to strictly follow the sequences ~\cite{pineles2014accuracy,mondol2015harmony,mondol2020hawad,samyoun2021iwash, wang2022handwashing, nagar2022hand, ma2023reducing, xu2024washring, asif2024resmfuse}. However, people always wash their hands in different gesture sequences. UWash is adaptable and can assess handwashing procedures regardless of the sequence of gestures used.

    \item \textbf{Detailed duration for each gesture.} 
   Most existing methods merely count the number of handwashing occurrences or measure the total duration of the process~\cite{pineles2014accuracy,samyoun2021iwash,mondol2020hawad,wang2020accurate, nagar2022hand, ma2023reducing, shrimali2023novel, asif2024resmfuse}. However, different gestures target different parts of the hands, making the duration of each specific gesture a crucial metric for assessing handwashing quality. UWash provides this information, ensuring a comprehensive evaluation of handwashing practices.
    
    \item \textbf{Instructive with scoring.} UWash can score the handwashing quality to guide users to improve their handwashing techniques. Since the WHO only provides an overall handwashing recommendation time of 40-60 seconds without specifying the duration for each stage of handwashing, we reviewed approximately 60 YouTube videos featuring expert handwashing demonstrations. From these, we carefully selected 12 videos and recorded the duration of each handwashing stage. We then averaged these durations to establish a gold standard for stage durations, which serves as the basis for our scoring, described in Sec.~\ref{sec:score}.
    
\end{itemize}
To our best knowledge, our UWash is the first work that provides all these features to meet the requirements of daily handwashing assessment as compared in Table~\ref{tab:related_work}.

\begin{table*}[t]
\centering
\caption{This table compares different handwashing assessment solutions based on key requirements, such as flexibility, sensor independence, and start/end detection. UWash uniquely meets all these requirements, setting it apart from other methods.}
\footnotesize
\begin{tabular}{lccccccc}
\toprule
\textbf{Work} & \textbf{no fixed position}  & \textbf{no extra sensor} & \textbf{start/end} & \textbf{random sequence} & \textbf{gesture duration} & \textbf{scoring} \\ 
\midrule
RGD camera, 2011~\cite{llorca2011vision}        & \XSolidBrush                      & \Checkmark                                    & \Checkmark                             & \Checkmark                       & \Checkmark                         &  \XSolidBrush             \\
RFID, 2014~\cite{pineles2014accuracy}          &  \XSolidBrush                        & \Checkmark                                       &  \XSolidBrush                             &  \XSolidBrush                       &  \XSolidBrush                          &  \XSolidBrush             \\
Harmony, 2015~\cite{mondol2015harmony}       &  \XSolidBrush                         &  \XSolidBrush                        &  \XSolidBrush                             &  \XSolidBrush                       & \Checkmark                          &  \XSolidBrush             \\
Depth camera, 2016~\cite{zhong2016washindepth}  &  \XSolidBrush                          &  \XSolidBrush                                &  \XSolidBrush                             & \Checkmark                       & \Checkmark                          &  \XSolidBrush             \\
WristWash, 2018~\cite{li2018wristwash}     & \Checkmark                          & \Checkmark                                       & \Checkmark                             & \Checkmark                       & \Checkmark                          &  \XSolidBrush             \\
RFWash, 2020~\cite{khamis2020rfwash}        &  \XSolidBrush                          & \Checkmark                                       &  \Checkmark                             & \Checkmark                       & \Checkmark                          &  \XSolidBrush             \\
HAWAD, 2020~\cite{mondol2020hawad}         & \Checkmark                          & \Checkmark                                     & \Checkmark                             &  \XSolidBrush                       &  \XSolidBrush                          &  \XSolidBrush             \\
Armband, 2020~\cite{wang2020accurate}       & \Checkmark                          & \Checkmark                                        & \Checkmark                             & \Checkmark                       &  \XSolidBrush                         &  \XSolidBrush             \\
iWash, 2021~\cite{samyoun2021iwash}         &  \XSolidBrush                          &  \XSolidBrush                         &  \XSolidBrush                             &  \XSolidBrush                       &  \XSolidBrush                          &  \XSolidBrush             \\
AWash, 2021~\cite{cao2021awash} &  \XSolidBrush                          &  \XSolidBrush                         & \Checkmark                              & \Checkmark                        & \Checkmark                           &  \XSolidBrush             \\

DLASS, 2022~\cite{wang2022handwashing} & \XSolidBrush & \Checkmark & \Checkmark & \XSolidBrush & \Checkmark & \XSolidBrush \\
Vision-based, 2022~\cite{nagar2022hand} & \XSolidBrush & \Checkmark & \Checkmark & \XSolidBrush & \XSolidBrush & \XSolidBrush \\
Smartwatch, 2022~\cite{lattanzi2022unstructured} & \Checkmark & \Checkmark & \Checkmark & \Checkmark & \XSolidBrush & \XSolidBrush \\
ALPHA HW, 2023~\cite{ma2023reducing} & \XSolidBrush & \Checkmark & \Checkmark & \XSolidBrush & \XSolidBrush & \XSolidBrush \\
CareHAI, 2023~\cite{shrimali2023novel} & \XSolidBrush & \XSolidBrush & \Checkmark & \Checkmark & \XSolidBrush & \XSolidBrush \\
WashRing, 2024~\cite{xu2024washring} & \Checkmark & \Checkmark & \Checkmark & \XSolidBrush & \XSolidBrush & \XSolidBrush \\
ResMFuse-Net, 2024~\cite{asif2024resmfuse} & \XSolidBrush & \Checkmark & \Checkmark & \XSolidBrush & \XSolidBrush & \XSolidBrush \\

\midrule
UWash~(ours)        & \Checkmark                          & \Checkmark                                      & \Checkmark                             & \Checkmark                       & \Checkmark                          & \Checkmark   \\
\bottomrule
\end{tabular}
\label{tab:related_work}
\end{table*}

Though UWash provides a variety of features and multiple functions, our novel methods make it simple. Generally, there are two main schemas for continuous gesture recognition with the time-serial readings. (1) Top-down. This schema first segments the readings into slices then applies gesture recognition on each slice. However, segmenting two contiguous gestures through the readings is challenging and always results in errors as explained in~\cite{khamis2020rfwash}. One strategy to relieve this challenge is to set a pause gesture between two contiguous gestures to make the corresponding readings easy to segment~\cite{ding2015femo}. However, this strategy will raise great inconvenience to users for handwashing assessment. (2) Bottom-up. This schema leverages a sliding window. All samples in the window would be categorized as the same gesture with time-series classification methods such as Hidden Markov Models~\cite{li2018wristwash}, Dynamic Time Warping~\cite{akyazi2017smokewatch}, Support Vector Machine~\cite{lee2016standalone}, Recurrent Neural Networks~\cite{hou2019signspeaker}, etc. As the window slides, entire readings will be recognized. This schema bypasses the segmenting problem in the top-down schema. However, this schema has an intrinsic false classification when the window spans readings of contiguous gestures.

We apply the bottom-up schema in UWash for it bypasses segmenting problem in the top-down schema. To tackle the intrinsic false classification of the bottom-up schema, we no longer conduct gesture recognition over the window, but conduct gesture semantic segmentation over the window. Semantic segmentation in the computer vision community is to align every pixel in an image with an object category, aka pixel-wise classification. In UWash, we adapt U-Net~\cite{ronneberger2015u,wang2023u}, a well-known semantic segmentation network for medical images, to align every sampling point (sample) in a sliding window with a gesture category. As the window slides, every sample in the time-serial IMU readings will have gesture alignment for itself, aka sample-wise gesture classification. To facilitate the gesture semantic segmentation results, Pyramid Pooling Module~\cite{zhao2017pyramid} and channel-wise attention~\cite{hu2018squeeze} are applied to the U-Net. Fig.~\ref{fig:overview} shows an example of gesture semantic segmentation on a handwashing procedure. With the estimated procedure segmentation, UWash is able to detect the start and end of handwashing, estimate the duration of each gesture, and score each gesture as well as the entire procedure according to the estimated duration. 

The technical novelty of our UWash lies in the synergy achieved by tackling the intrinsic problem in the bottom-up schema with gesture semantic segmentation strategy, picking the right techniques such as U-Net~\cite{ronneberger2015u}, Pyramid Pooling Module~\cite{zhao2017pyramid} and channel-wise attention~\cite{hu2018squeeze}, and proposing a Dual-Branch U-Net framework for the two modal inputs from accelerometers and gyroscopes of smartwatches. In conclusion, the proposed methods provide five main features as follows.
\begin{itemize}
    \item \textbf{Lightweight.} UWash is lightweight, only 496KB.
    
    \item \textbf{Simple.} UWash addresses the intrinsic problem in the bottom-up schema with the simple idea, i.e., gesture semantic segmentation. Besides, to make U-Net suitable for time-serial IMU readings, we simply replace its 2D operations, e.g., convolution and pooling, with the 1D versions.   
    
    \item \textbf{Effective.} Extensive evaluation results show UWash performs well in multiple settings, e.g., user-dependent, cross-user, cross-location, and cross-time.
    
    \item \textbf{Fine-grained.} UWash provides sample-wise classification results, which is fine-grained compared to conventional top-down schema (slice-wise classification) and bottom-up schema~(window-wise classification). 
    
    \item \textbf{One model for multiple tasks.} UWash models can automatically detect the start and end of the handwashing procedure, estimate the duration of each gesture, and score each gesture as well as the entire procedure.
\end{itemize}

The contributions of this paper can be summarized in the following four aspects.

\textbf{(1)} We propose UWash to automatically assess handwashing techniques for people's daily use, which is the first work that provides all features listed in Table~\ref{tab:related_work}. 

\textbf{(2)} We novelly regard the handwashing assessment task as the semantic segmentation task to bypass the intrinsic problem in the bottom-up time-serial classification schema, and design a U-Net variant to achieve it well.

\textbf{(3)}  We propose a simple approach to obtain a standard of the expected duration of each handwashing gesture following WHO guidelines. With the standard, UWash is the first work that can score the handwashing quality to guide users to improve their handwashing techniques.

\textbf{(4)} We collect a dataset with 51 subjects and 5 locations, and conduct an extensive evaluation in settings of user-dependent, cross-user, and cross-location. Besides, we collect a new dataset 9 months later in a hospital to demonstrate the cross-time performance. All datasets and codes have been released for research use.

\section{Related Work}~\label{sec:related_work}
Hand hygiene has been crucial to preventing healthcare-associated infections in hospitals. Human mandatory audits are applied to improve the healthcare workers' compliance with the WHO guidelines. However, this approach is labor-intensive, time-consuming, and costly. Thus automatic handwashing monitoring systems are proposed~\cite{kinsella2007electronic, pineles2014accuracy, rabeek2014reliable, edmond2010successful, llorca2011vision, zhong2016washindepth, haque2017towards, li2018wristwash, khamis2020rfwash, samyoun2021iwash} to facilitate healthcare workers' adherence. Among them, sensors such as alcohol sensors~\cite{edmond2010successful}, pressure sensors~\cite{kinsella2007electronic}, ultrasonic hotspots~\cite{rabeek2014reliable}, and RFID~\cite{pineles2014accuracy} are embedded into the dispensers to simply count the handwashing. RGB cameras~\cite{llorca2011vision, wang2022handwashing, nagar2022hand, ma2023reducing, shrimali2023novel, asif2024resmfuse}, depth cameras~\cite{zhong2016washindepth, haque2017towards}, and mmWave devices~\cite{khamis2020rfwash} are placed on the wall to estimate fine-grained handwashing procedures. Since sensors or devices in these works are required to be deployed next to the handwashing sinks in hospitals, they are not suitable for people's daily use.

Several wearable and infrastructure-based systems have been proposed for handwashing assessment. WristWash~\cite{li2018wristwash} utilizes wrist-worn devices to evaluate handwashing through clip-wise motion recognition but is constrained by predefined sequences, reducing its adaptability to diverse user behaviors. Similarly, iWash~\cite{samyoun2021iwash} relies on smartwatch activation via user interaction, which may discourage frequent use in daily settings.WashRing~\cite{xu2024washring} introduced a hand hygiene monitoring system based on smart ring devices. Although smart rings have the advantages of being compact and unobtrusive to wear, they may interfere with the handwashing process when worn on the fingers.In addition to wearable devices, environmental sensor-based methods such as AWash~\cite{cao2021awash} and RFWash~\cite{khamis2020rfwash} have also been explored. AWash~\cite{cao2021awash} employs smart faucets or soap dispensers to detect handwashing onset and offset, but its reliance on specific infrastructure limits deployment flexibility. RFWash~\cite{khamis2020rfwash} uses RFID tracking for hand hygiene compliance but lacks the granularity to analyze individual gestures or provide detailed feedback. Besides requirements in user experiences, we believe if the handwashing monitoring system can score every individual handwashing gesture and the entire procedure following WHO guidelines, it will help people to improve their handwashing techniques following the reported scores. Thus we propose UWash to achieve this.

In Table~\ref{tab:related_work}, we compare UWash with some representative work. Unlike infrastructure-based methods like AWash~\cite{cao2021awash} and RFWash~\cite{khamis2020rfwash}, which require external sensors or specific setups, UWash relies solely on smartwatch IMU sensors, making it more versatile and accessible for daily use. Compared to wearable-based systems like WristWash~\cite{li2018wristwash} and WashRing~\cite{xu2024washring}, UWash supports random gesture sequences and avoids the usability challenges posed by predefined patterns or potential interference during handwashing.
Moreover, UWash is the first work with the capability to score the handwashing quality, which evaluates the quality of each gesture and the entire procedure based on WHO guidelines. This capability not only ensures compliance but also provides actionable feedback to users, promoting better hygiene practices. Additionally, UWash automates the detection of handwashing start and end events, eliminating the need for manual activation or external triggers. These advantages make UWash a comprehensive and practical solution for hand hygiene monitoring.

\begin{figure*}[t]
    \centering
    \includegraphics[width=1\linewidth]{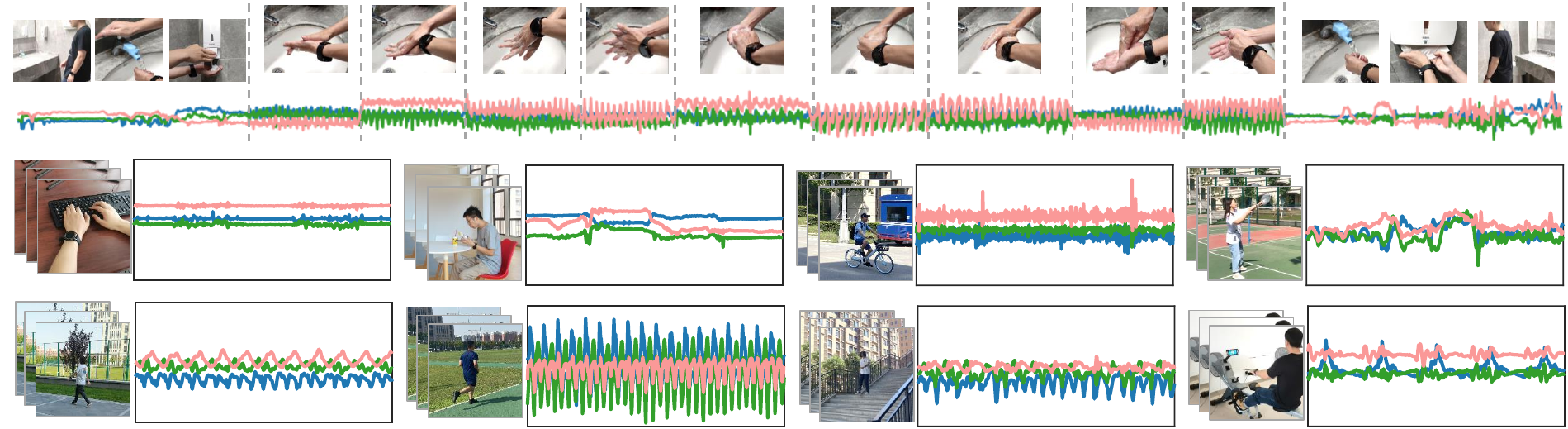}
    \caption{Uwash can automatically detect the event of handwashing, not requiring to work along with Bluetooth sensors in dispensers~\cite{mondol2015harmony} or to awaken the monitoring system manually~\cite{samyoun2021iwash}. We attribute this ability partly to the unique characteristics explained in Section~\ref{sec:character}, i.e., successive, periodical, and being with expectable pre-activities and post-activities. 
 }
    \label{fig:examples}
\end{figure*}

\section{Handwashing Analysis }~\label{sec:analysis}

% \vspace{-28pt}
\subsection{Features of Handwashing}~\label{sec:character}
As mentioned above, UWash can automatically detect the handwashing activity, not requiring to work along with additional sensors~\cite{mondol2015harmony,samyoun2021iwash,zhong2016washindepth} or to be activated manually~\cite{samyoun2021iwash}.
In addition to the carefully-designed algorithm that we will describe in detail later in Section~\ref{sec:method}, we also attribute this ability to the unique features of the handwashing activity. With the visualization study in Fig.~\ref{fig:examples}, we conclude three unique features as follows. 

(1) Multiple successive gesture stages. When washing hands, people always conduct several successive gestures to clean the palms, back of hands, fingers, wrists, etc., respectively. We visualize one example recorded by the accelerometers in Fig.~\ref{fig:examples}, from which we can find distinguishable signal patterns of multiple successive stages. 

(2) Periodic motion under each gesture. When cleaning a part of hands, people always perform periodic back and forth motions. Besides, the periodic records in multiple successive stages are very different, as shown in Fig.~\ref{fig:examples}. In contrast, we show records of four non-periodic daily activities, i.e., keystroking, eating, cycling, and playing badminton, as well as four periodic activities, i.e., walking, running, walking downstairs, and rowing. These daily activities are either single-stage, not periodic, or neither. Therefore, handwashing procedures can be detected automatically. 

(3) Expectable pre-activities and post-activities. The pre-activities of handwashing procedures always include walking to a sink, wetting hands, applying soap, etc. The post-activities of handwashing procedures always include drying hands with a towel, dropping the towel, walking out of the restroom, etc. These activities are priors that largely promote our start/end time detection.

Since handwashing is a continuous, multi-stage process, where each stage involves repeated actions, while some daily activities may be misclassified as belonging to a particular handwashing phase, the need for specific criteria, such as a period of duration, multiple stages, and repeated actions within each stage, can help filter out these misclassifications.

 \begin{figure}[t]
    \centering
    \includegraphics[width=1\linewidth]{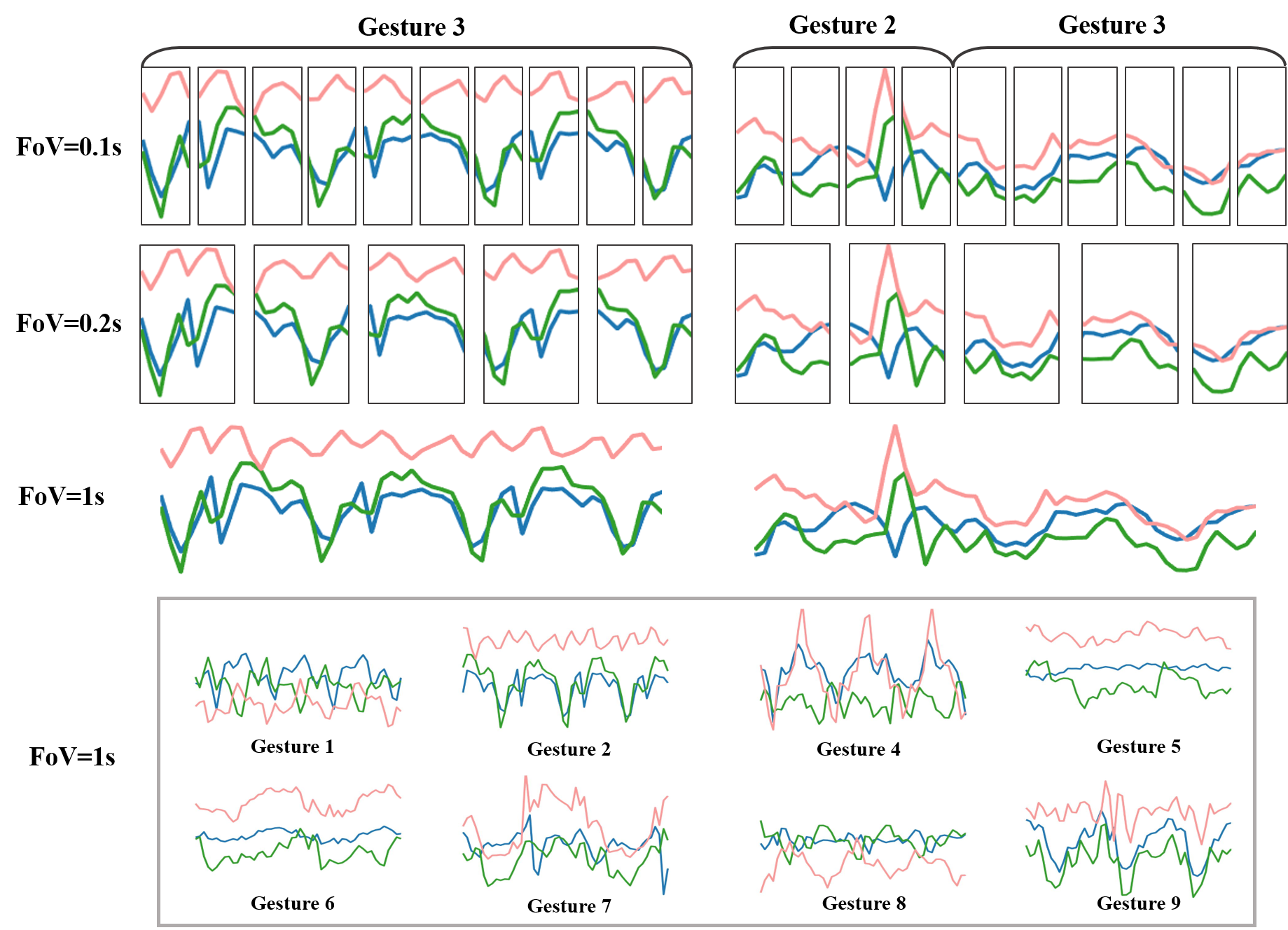}
    \caption{Handwashing patterns emerge in a larger field of view (left). However, a larger field of view may lead to a large prediction error under window-grained prediction when handwashing gestures switch (right). Therefore, we apply the semantic segmentation method for sample-wise prediction in the window to bypass the dilemma.}
    \label{fig:dilemma}
\end{figure}

\subsection{Dilemma on Field of View}~\label{sec:dilemma}
The handwashing procedures vary from person to person, even from time to time for the same person, leading to diverse handwashing motion sequences. Conventional action recognition approaches use sliding windows with a data-dependent stride to crop clips from sequences. Then gesture classification is conducted clip-by-clip. To facilitate understanding, here we call the size of sliding windows the Field of View~(FoV), which represents the receptive field that handwashing gesture recognition models can see at one time. 

Selecting FoV in the conventional FoV-wise recognition schema is a work of dilemma. As shown in Fig.~\ref{fig:dilemma}, the left three sub-figures are an example of recorded accelerometer sequences with the FoV of 0.1~\textit{seconds}, 0.2~\textit{seconds}, and 1~\textit{second}, respectively. Let us see the upper-left first, we can hardly tell the gesture category from one single sequence, for these 10 sequences are quite different. If we take a twice-wider FoV of 0.2~\textit{seconds}, the tendency of the sequences becomes much clear, e.g., sequences in the 1st and 4th windows are similar; and those in the 2nd and 5th windows are similar. However, the blue/green curves in the 1st/4th window increase, while the blue/green curves in the 2nd/5th window decrease. This leads to an ambiguity in the handwashing gesture recognition. Furthermore, if we take an even wider FoV of 1~\textit{second}, the unified and periodical patterns in the sequence finally emerge, with which we can easily make an accurate handwashing gesture recognition. 

{\textbf{Pro.} } Larger FoV facilitates handwashing gesture recognition.

However, a larger FoV may also cause a larger error. As shown in Fig.~\ref{fig:dilemma}, the right three sub-figures demonstrate a procedure of gesture switching from gesture 2~(G2) to gesture 3~(G3), where G2 and G3 occupy the duration in 40\% and 60\% respectively. As G3 occupies the dominant duration, gesture recognition models tend to classify the sequences as G3,  resulting in a recognition error of 40\%.

{\textbf{Con.}} Larger FoV may lead to larger recognition errors. 

To relieve the dilemma on the FoV, one could search for a dataset-specific FoV or propose a dynamic FoV algorithm. This may be over-designed with complex strategies. It is unclear how we can ensure every gesture duration is perfectly divisible by FoVs. Our solution to the dilemma is to borrow the semantic segmentation schema~\cite{long2015fully} for the handwashing gesture recognition task. Instead of outputting one single gesture category for one entire FoV, the semantic segmentation schema predicts the gesture category for every single sample in an FoV. This schema takes the advantage of a large FoV and makes fine-grained recognition for every moment in the FoV, bypassing the dilemma on FoVs naturally. We will explain our methods in detail in Section~\ref{sec:method}.

\begin{figure*}[t]
    \centering
    \includegraphics[width=1\linewidth]{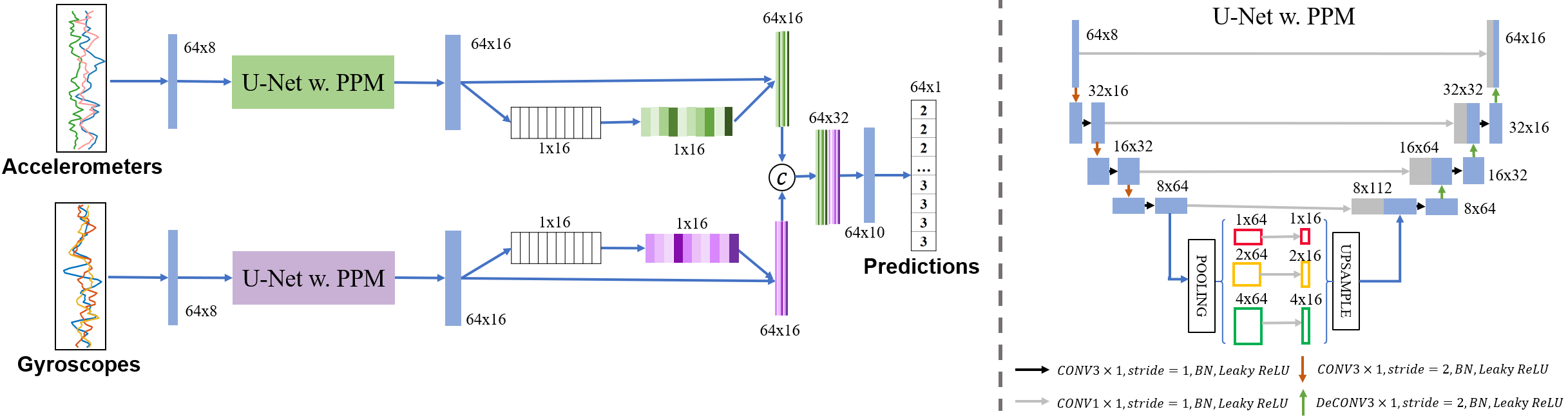}
    \caption{Deep Network Architecture of UWash. The dual-branch U-Nets~\cite{ronneberger2015u} take data from two modality sensors, i.e., accelerometers and gyroscopes, as inputs, respectively. Feature maps from two branches are further concatenated in high-level layers for sample-wise gesture recognition. Since accelerometers and gyroscopes have different modality measures of handwashing gestures, before concatenating, the squeeze-and-excitation modules~\cite{hu2018squeeze} automatically normalize their feature maps.}
    \label{fig:network}
\end{figure*}

\section{Methods}~\label{sec:method}
Before going into details of the methods, we define our task with symbols. We use $N$ to represent the length of the FoV, which is a pre-defined constant. We use $K$ to represent the size of the training dataset. We use $A$ and $G$ to represent data of the accelerometers and gyroscopes, respectively. We use $Y$ for the sample-wise gesture annotations. With these symbols, the training dataset is $\mathcal{D}= \{A_i^j,G_i^j; Y_i^j ~|~ i\in 1,2,..,K; j\in 1,2,...,N\}$. Please note that accelerometers and gyroscopes both record data in 3 dimensions, which are not explicitly shown for brevity. We further simplify the notation of the training dataset as $\mathcal{D}=\{A,G; Y\}$.

Our goal is to propose a machine learning model $\mathcal{M}$ that takes $A$ and $G$ as inputs, and outputs sample-wise gesture recognition results, $Y^*$. We conclude this goal with Equation~\ref{eq:target}.
\begin{equation}~\label{eq:target}
\centering
    \mathcal{M} =  \arg \min_\mathcal{M} \left \|  \mathcal{M}(A,G) \rightarrow Y^*, Y   \right \|
\end{equation}
where $\left \| \cdot \right \|$ is for the operator to compute distances between the model's outputs~($Y^*$) and annotations~($Y$).

\subsection{Deep Learning Model}~\label{sec:deep_model}
U-Net~\cite{ronneberger2015u} is a pixel-wise classification network architecture widely used in computer vision tasks of visual semantic segmentation. Temporal U-Net~\cite{wang2019temporal} replaces 2-dimensional convolutions in U-Net with 1-dimensional convolutions to conduct sample-wise class alignment on 1-dimensional time-serial data. The U-Net clusters can learn features from data in the wider FoVs with the increase of the convolutional layers~(please refer to Section~\ref{sec:dilemma} for the definition of the FoVs). Further, learned features in large FoVs~ (at the high-level layers) and small FoVs~(at the low-level layers) are combined with skip connections to promote semantic segmentation performance. Considering these good characters on segmentation, we apply Temporal U-Net with several modifications to achieve our task of the sample-wise handwashing gesture recognition. The network architecture is shown in Fig.~\ref{fig:network}. Next, we explain the reason why we propose these modifications.

\textbf{(1) Two-stream.} Accelerometers and gyroscopes of smartwatches measure linear accelerations and angular accelerations respectively, describing two physical quantities in different scales. To properly leverage data of the two modalities, we have to do data normalization before merging them for the later task. The classic Gaussian normalization method requires the mean and the variance of the whole training dataset, which is highly dataset-dependent and sensitive to extremums.
To conduct automatic modality normalization, we feed raw accelerometer data and raw gyroscope data into the two streams of U-Nets, respectively. We also apply Batch Normalization~\cite{ioffe2015batch} and Leaky Rectified Linear Unit~\cite{maas2013rectifier} to facilitate stable training and fast convergence. 

\textbf{(2) Squeeze-and-Excitation Module~\cite{hu2018squeeze}.} Though features learned from the dual-branch U-Nets are considered to be normalized, we still believe their contributions to the final gesture recognition are always not equal. The reason is concluded from the observation that accelerometers and gyroscopes always have different sensitivity when one conducts a specific handwashing gesture. For example, as shown in Fig.~\ref{fig:overview}, accelerometers are more sensitive than gyroscopes for G1, while gyroscopes are more sensitive for G6. To re-weight the contribution of handwashing gesture recognition of the dual-modal sensors, we apply Squeeze-and-Excitation Modules to the learned features. After that, we concatenate the re-weighted features along the channel dimension for gesture recognition.

\textbf{(3) Pyramid Pooling Module~(PPM)~\cite{zhao2017pyramid}.} PPM is proven to be efficient in harvesting features across different FoVs~\cite{zhao2017pyramid}. Therefore, we apply it in the middle of U-Nets, shown in Fig.~\ref{fig:network}, to harvest IMUs features of different FoVs in our task. Before being fed into PPM, feature maps are with the size of $8\times64$, where 8 and 64 represent temporal dimension and channel dimension, respectively. We use three average pooling operations with window/stride sizes of 8, 4, and 2 on the feature maps, generating three outputs with the size of $1\times64$, $2\times64$, and $4\times64$, respectively. Before concatenating three pooling outputs with the input feature maps along the channel dimension, we use convolution layers with the kernel size of $1\times 1$ to reduce the channel to be of 16, for the channel information balance between the 3 pooling outputs and the input feature maps. Further, we upsample the pooling outputs and concatenate them with the input feature maps, outputting feature maps with the size of $8\times 112$.

\textbf{(4) Input length of 64.} We have two reasons to set the input clip length as 64~(FoV of 64). First, as shown in Fig.~\ref{fig:dilemma}, FoV with the size of 1~\textit{second}~is sufficient to discriminate gestures. For the sampling rate of the smartwatch is 50Hz, the nearest number is 64~(with an integer power of 2). Second, since larger inputs lead to larger models, we set the input length to be 64 instead of 128 or larger for the future deployment on edge devices. Actually, the UWash models are quite lightweight, only 496KB without any model compression, making it possible to deploy UWash in any smartwatches in the future. 

We use Pytorch 1.9.0 to implement the network. The initial learning rate is 0.001. The batch size is 16k. We use the cross-entropy function to compute losses and Adam~\cite{kingma2014adam} to optimize the network. We train the network for 500 epochs.

\begin{figure}[t]
    \centering
    \includegraphics[width=1\linewidth]{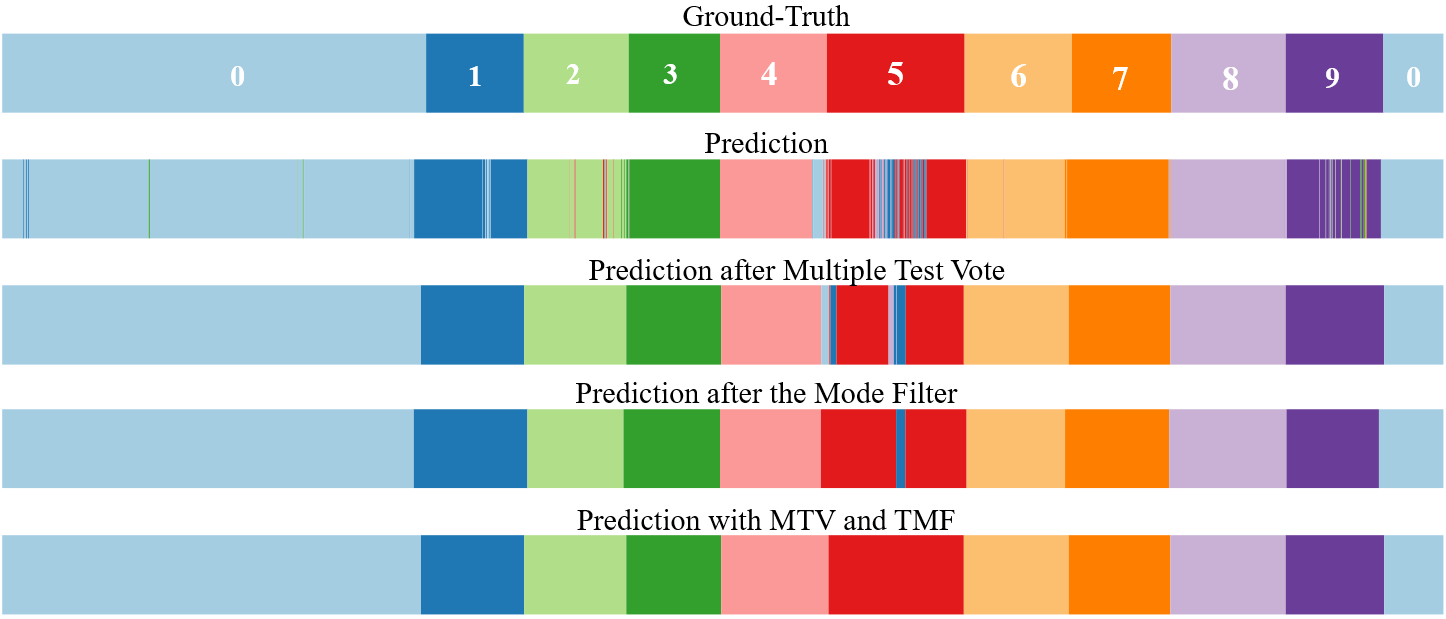}
    \caption{A post smoothing example. UWash can do sample-wise recognition well directly, while jitter errors cause the majority of the false recognition. To reduce these errors, we propose post smoothing methods including multiple test voting and the mode filter, which are simple but work effectively.}
    \label{fig:smooth}
\end{figure}

\begin{figure}[t]
    \centering
    \includegraphics[width=0.8\linewidth]{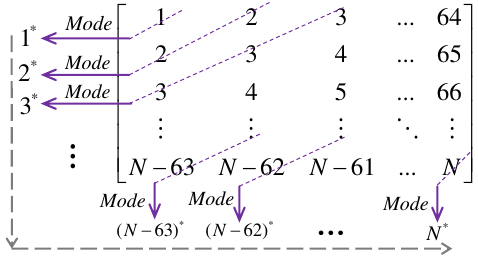}
    \caption{Multiple Test Voting. We conduct gesture recognition over the test time series via the FoV of 64 with the~\textit{stride of 1}, which results in multiple outputs on each sampling point. For each sampling point within the time series, we take the mode of all its outputs as its final recognition result.}
    \label{fig:mtv}
\end{figure}

\subsection{Post Smoothing Methods}\label{sec:smooth}
In testing, given a test time series with the length of $N$, we conduct gesture recognition with the FoV of 64 and the stride of 64. We show a raw recognition output in Fig.~\ref{fig:smooth}, which shows some jitter errors due to random false classification. Thus, we further apply two post-smoothing methods to the raw recognition outputs: 

(1) Multiple Test Voting~(MTV). We conduct gesture recognition over the test time series via the FoV of 64 with the~\textit{stride of 1}, which results in multiple outputs on each sampling point, as shown in Fig.~\ref{fig:mtv}. For each sampling point within the time series, we take the output that appears most often, i.e., the mode, of all its outputs as its final recognition result.

(2) The Mode Filter~(TMF). For each sample, we use the mode of outputs on its nearest 128 samples as its final recognition result. We call this method the Mode Filter with a window size of 128 and stride of 1.  

A post-smoothed example is visualized in Fig.~\ref{fig:smooth}, which shows that MTV and TMF can effectively reduce jitter errors and improve sample-wise gesture recognition. The numerical comparison results will be reported in Section~\ref{sec:evaluation}.

\begin{table}[t]
\centering
\caption{Professional Duration of Each Gesture. We select 12 videos to record the recommended duration of each gesture in these videos. For each gesture, the maximums (highlighted in \textcolor{red}{\sout{red}}) and minimums (highlighted in \textcolor{blue}{\sout{blue}}) are excluded to mitigate outlier effects. The final reference duration for each gesture is the average of the remaining 10 videos.}
\label{tab:guidelines}
 \setlength{\tabcolsep}{3pt} 
\begin{tabular}{cccccccccccccc}
\toprule
\textbf{No.} & \textbf{1} & \textbf{2} & \textbf{3} & \textbf{4} & \textbf{5} & \textbf{6} & \textbf{7}  & \textbf{8} & \textbf{9} & \textbf{10} & \textbf{11} & \textbf{12} & \textbf{Avg.} \\ \midrule
\textbf{G1}  & \textcolor{blue}{\sout{3}}          & 4          & 4          & 4          & 5          & 6          & 4         & 5          & 3          & \textcolor{red}{\sout{9}}           & 7           & 7           & 4.9          \\ 
\textbf{G2}  & \textcolor{blue}{\sout{1.5}}        & 3.5        & 3          & 4.5        & 3          & 3.5        & 4.5       & 3.5        & 3.5        & \textcolor{red}{\sout{6.5}}         & 4.5         & 3           & 3.65         \\ 
\textbf{G3}  & \textcolor{blue}{\sout{1.5}}        & 3.5        & 3          & 4.5        & 3          & 3.5        & 4.5       & 3.5        & 3.5        & \textcolor{red}{\sout{6.5}}         & 4.5         & 3           & 3.65         \\ 
\textbf{G4}  & 4          & 4          & 3          & 5          & 6          & 5          & 6        & 6         & \textcolor{red}{\sout{11}}         & 10          & 5           & \textcolor{blue}{\sout{2}}           & 5.4          \\ 
\textbf{G5}  & \textcolor{blue}{\sout{2}}          & 4          & 3          & 5          & 5.5       & 3.5        & 5        & 3.5     & 3.5        & \textcolor{red}{\sout{8.5}}        & 4           & 3           & 4            \\ 
\textbf{G6}  & \textcolor{blue}{\sout{2}}          & 3          & 2          & \textcolor{red}{\sout{5.5}}        & 4.5        & 2.5        & 4      & 3.5      & 3.5        & 4           & 5           & 2.5         & 3.45         \\ 
\textbf{G7}  & \textcolor{blue}{\sout{2}}          & 3          & 2          & \textcolor{red}{\sout{5.5}}        & 4.5        & 2.5        & 4      & 3.5      & 3.5        & 4           & 5           & 2.5         & 3.45         \\ 
\textbf{G8}  & 3          & 3          & \textcolor{blue}{\sout{2.5}}        & 4.5        & 5.5        & 3.5        & \textcolor{red}{\sout{6.5}}     & 3        & 4          & 6           & 5           & 3.5         & 4.1          \\ 
\textbf{G9}  & 3          & 3          & \textcolor{blue}{\sout{2.5}}        & 4.5        & 5.5        & 3.5        & \textcolor{red}{\sout{6.5}}    & 3       & 4          & 6           & 5           & 3.5         & 4.1          \\ \bottomrule
\end{tabular}
\end{table}

\subsection{Handwashing Scoring}~\label{sec:score}
Scoring the quality of handwashing procedures is subjective, which requires expertise in the type of gestures, the completion of gestures, the duration of gestures, etc. In this paper, we directly evaluate the UWash quality with respect to the WHO guidelines. To obtain the guideline-recommended duration of each gesture, we collected 60+ handwashing videos that describe WHO guidelines and choose 12 of them as references, ignoring those with slow play, fast play, over-detailed explanation, etc. Urls of seleted videos are as follows.\\
\noindent (1) \url{https://www.youtube.com/watch?v=qo7Q_wm2Vec}\\
(2) \url{https://www.youtube.com/watch?v=IisgnbMfKvI}\\
(3) \url{https://www.youtube.com/watch?v=0FLQ-EpQ6PM}\\
(4) \url{https://www.youtube.com/watch?v=jXqDAfeUFBg}\\
(5) \url{https://www.youtube.com/watch?v=6JrEeR5OXiE}\\
(6) \url{https://www.youtube.com/watch?v=TClRYmtqClM}\\
(7) \url{https://www.youtube.com/watch?v=a9CMtzymZTg}\\
(8) \url{https://www.youtube.com/watch?v=3PmVJQUCm4E}\\
(9) \url{https://www.youtube.com/watch?v=hhKlyoVsbOY}\\
(10) \url{https://www.youtube.com/watch?v=3KjUaibd4gg}\\
(11) \url{https://www.youtube.com/watch?v=4CcGLoYrIPU}\\
(12) \url{https://www.youtube.com/watch?v=YiChdJ_os3Q&t=53s}

Table~\ref{tab:guidelines} shows the recommended duration of each gesture in these videos. In addition, we remove the maximum and minimum of each gesture and compute the average as the professional handwashing duration, denoted as $D_i^p, i\in[1,2,...,9]$.
We have two empirical assumptions. (1) Since each gesture emphasizes cleaning one part of the hands, we assume each gesture is equally important in handwashing. (2) We assume the quality of cleaning of each gesture increases linearly with its duration, and the perfect quality is reached and saturated when the duration is equal to or greater than the professional duration. Given these assumptions, we score handwashing with Equation~\ref{eq:score}.

\begin{equation}\label{eq:score}
    Score= \frac{100}{9} \sum_{i=1}^{9}min(1,\frac{D_i^e}{D_i^p})
\end{equation}
where $\frac{100}{9}$  is the best score of each gesture, to match the first assumption; $D_i^e$ represents the estimated duration of the $i$-th gesture; $min(1,\frac{D_i^e}{D_i^p})$ matches the second assumption. Finally, the $Score$ of a handwashing procedure is the sum of estimated scores of all nine gestures. Note that the above two assumptions only consider handwashing duration and do not incorporate critical quality determinants of hygiene efficacy, e.g., soap coverage and scrubbing vigor.

obscure differences in gesture quality (e.g., whether soap was properly applied or
hands fully covered). 

It is important to highlight that the gold standard for evaluating actual handwashing quality involves professional personnel conducting video observations. This process may consider multiple factors, including the posture of each gesture, the thoroughness of the gestures, and the water flow rate at the sink. Our two assumptions are simplified models for assessing handwashing quality with low cost.

Please note that we have taken applying gels into consideration since it is challenging to quantify the scores of various types of sanitization gels (e.g., spray, gel, or foam) and the different ways users apply them (e.g., pressing with fingers or palms).

\section{Evaluation}~\label{sec:evaluation}

\vspace{-20pt}
\subsection{Data Acquisition}~\label{sec:data}
This study was approved by the Medical Ethics Committee of the Second Affiliated Hospital of Xi'an Jiaotong University, Xi'an China.

We use the Samsung Gear Sport smartwatches and collect data from motion sensors and corresponding timestamps following~\cite{fomichev2019perils}. To increase the diversity of external conditions such as the type of hydrants, sinks, dispensers, etc., we collect data at five buildings on campus, i.e., a teaching hall, a laboratory hall, a cafeteria, a dormitory, and a library. At each building, we randomly recruit 10 passersby~(11 at the laboratory hall) as participants and train them to wash their hands following WHO guidelines.

To act as the daily handwashing procedure, participants were asked to conduct activities including walking to the sink, washing hands, and walking out of the restroom, while other activities such as wetting hands with water, applying soap, and drying hands with a towel are not mandatory, depending on their behaviors. We denote gestures in WHO guidelines as category 1 to category 9, and all other activities as category 0. Every participant repeats the procedure 5 times, which is a tolerable number and would not cause any hand discomfort. Along with the sensor data, we also use codes from~\cite{wang2019person} to record videos on the participants' hands and corresponding timestamps. Five labeling workers view the synchronized video streams to label gestures on motion sensory data collected at each building respectively. Further, we segment motion sensory records with the FoV of 64 and the stride of 1, leading to a simple data augmentation of 63 times on the raw data, and resulting in a dataset of 804991 instances. To avoid ambiguity, we refer to data in an FoV as an~\textit{instance}, and data recorded at each sampling moment as a~\textit{sample}.

\begin{table*}[t]
\caption{Results on all participants. The results show that UWash performs well and can be enhanced with post-smoothing methods of multiple test voting~(MTV) and the mode filter~(TMF). Moreover, 
UWash is much more lightweight with $\sim$2.5\% parameters of ResNet-18-1D. Two-stream~(TS), Squeeze-and-Excitation Module~(SE), Pyramid Pooling Module~(PPM). 
$\uparrow$ means the larger the better. $\downarrow$ means the less the better.}
\centering
\label{tab:alluser}
\begin{tabular}{lccccc}
\toprule
   Method   & Accuracy~$\uparrow$ & mPrecison~$\uparrow$ & mRecall~$\uparrow$ & mF1~$\uparrow$ & Parameters $\downarrow$  \\ \midrule
     MobileNetV3-small~\cite{koonce2021mobilenetv3} & 82.95\% &80.50\%& 79.88\%& 0.80 & 3.056M (79\%)\\
      ResNet-18-1D~\cite{he2016deep} & 84.53\% &82.40\% &81.87\% & 0.82 & 3.851M (100\%)\\
      vanilla UNet~\cite{ronneberger2015u} & 82.25\% & 80.98\%&80.23\% & 0.81 &  0.048M (1.2\%)\\ 
      UNet+TS & 83.36\% & 81.96\% &81.43\% & 0.82& 0.098M (2.5\%)\\ 
       UNet+TS+SE & 84.16\% & 82.76\%& 81.87\%& 0.82&0.099M (2.5\%)\\ 
       UNet+TS+SE+PPM & 86.31\%        & 84.92\%        & 84.48\%      & 0.84  & 0.099M (2.5\%) \\
 UNet+TS+SE+PPM+MTV & 91.10\%        & 90.08\%        & 89.68\%      & 0.89  & 0.099M (2.5\%)\\
 UNet+TS+SE+PPM+TMF & 91.13\%        & 90.17\%       & 89.45\%      & 0.89 & 0.099M (2.5\%) \\
 UNet+TS+SE+PPM+MTV+TMF (UWash) & \textbf{92.27\%}        & \textbf{91.26\%}        & \textbf{90.85\%}      & \textbf{0.91} & 0.099M (2.5\%) \\ 
      \bottomrule
\end{tabular}
\end{table*}

\subsection{Evaluation Metrics} \label{sec:metrics}

\textbf{(1) Gesture recognition accuracy. } The accuracy is computed via Equation~\ref{eq:accuray}.
  \begin{equation}\label{eq:accuray}
     Accuracy = \frac{\sum_{p=1}^{51}\sum_{i=1}^{N_p} I( S_{p,i}^*==S_{p,i})}{\sum_{p=1}^{51}\sum_{i=1}^{N_p}}    
 \end{equation}
where $N_p$ represents the length of the test time-series of the $p$-th participant; $S_{p,i}$ and $S_{p,i}^*$ represent the ground-truth and the UWash output on the $i$-th sample of the $p$-th participant, respectively; $I( S_{p,i}^*==S_{p,i})$ outputs 1 if UWash recognizes correctly on $S_{p,i}$, otherwise 0.

\textbf{(2) Participant-wise accuracy.} For the $p$-th participant, we use Equation~\ref{eq:useraccuray} to compute the recognition accuracy of her/him.
 \begin{equation}\label{eq:useraccuray}
     Accuracy_p = \frac{\sum_{i=1}^{N_p} I( S_{p,i}^*==S_{p,i})}{N_p}    
 \end{equation}
 where symbols share the same meanings with Equation~\ref{eq:accuray}.

\textbf{(3)} We also apply precision, recall, and F1 to evaluate the gesture recognition performance.

\textbf{(4) Start/End Detection Error.} We use Equation~\ref{eq:errorstart} to compute the start time detection error. 
\begin{equation}\label{eq:errorstart}
   Error_s = \left | t_s^*-t_s\right |
\end{equation}
where $t_s^*$ and $t_s$ represent the detection and the ground-truth of the start time of a test handwashing procedure, respectively; $\left |\cdot\right |$ is to compute the absolute value. Similarly, we use $\left | t_e^*-t_e\right |$
to compute the detection error at the end time. 

\textbf{(5) Scoring Error.} UWash is the first work to score handwashing following the WHO guidelines. We use methods described in Section~\ref{sec:score} to compute handwashing scores on the test set.  We then use Equation~\ref{eq:error_scoring} to compute the scoring error. 
\begin{equation}\label{eq:error_scoring}
   Error_s = \left | \mathrm{score}^*-\mathrm{score}\right |
\end{equation}
where $\mathrm{score}^*$ and $\mathrm{score}$ represent the estimation and the ground-truth of handwashing scores, respectively.

\subsection{User-Dependent Results}\label{sec:results}
We first evaluate UWash on all participants under the user-dependent setting. For each participant, we use instances corresponding to the first 4 handwashing procedures as the training set, and the last ones as the test set, leading to the training and test set with instances of 643971 and 161020, respectively. In this paper, the training set and the test set have no overlap in all evaluations. 

 \textbf{(1) Overall Results.} We report overall results including accuracy, precision, recall, and F1 score in Table~\ref{tab:alluser}. The table shows that UWash can achieve the sample-wise classification accuracy of 86.31\% directly. With simple yet efficient post smoothing methods, i.e., multiple test voting and the mode filter~(described in Section~\ref{sec:smooth}), UWash can eventually achieve an accuracy of 92.27\%.

This is a 10-class classification task (9 handwashing gestures + 1 background gesture), so we have 10 precisions, 10 recalls, and 10 F1 scores. We report their means as mPrecision, mRecall, and mF1 in Table~\ref{tab:alluser}. Consistent with the accuracy, the results of these three metrics show that UWash performs well, and can be further enhanced with multiple test voting~(MTV) and the mode filter~(TMF). We also report the ablation study on our adapted methods, i.e., Two-Stream~(TS), Squeeze-and-Excitation~(SE) , and Pyramid Pooling Module~(PPM) in Table~\ref{tab:alluser}. The table shows that these methods can effectively improve the accuracy of one-stream vanilla UNet.

We further apply representative MobilenetV3-small~\cite{koonce2021mobilenetv3} and ResNet-18-1D~\cite{he2016deep} on the test clips to conduct clip-wise gesture classification. We take the clip-wise result as the result of all samples in the clip, and compute the sample-wise accuracy via Eq.~\ref{eq:accuray}. Table~\ref{tab:alluser} shows that UWash is $\sim$8-10\% higher than MobilenetV3-small and ResNet-18-1D. More importantly, MobilenetV3-small and ResNet-18-1D on our task are with 3.056M and 3.851M parameters respectively, while UWash is only with 0.099M parameters, $\sim$2.5\% of ResNet-18-1D, much more lightweight.

Table~\ref{tab:sota} compares our UWash with several recent work in view of accuracy, number of evaluated subjects, and granularity. UWash has been evaluated over the largest number of subjects, and achieves competitive accuracy with work that conducts clip-wise as well as sample-wise gesture classification. It is important to note that this comparison is only based on the granularity of the methods and the size of the datasets. Since the datasets are different, the differences in accuracy are not meaningful for reference.

\begin{table}[t]
\centering
\caption{Comparison with existing work in view of number of evaluated subject and granularity. Since the datasets are different, the differences in accuracy are not meaningful for reference }
 \setlength{\tabcolsep}{3pt} 
\begin{tabular}{lcccc}
\toprule
Work       & Accuracy     & \#Subject & Granularity        & Open-sourced       \\ \midrule
WristWash~\cite{li2018wristwash}  & 95\%         & 6         & clip-wise    &Not yet     \\ 
AWash~\cite{cao2021awash}      & 92.94\%      & 8         & clip-wise    & Not yet   \\ 
iWash~\cite{samyoun2021iwash}      & 92$\sim$98\% & 14        & clip-wise     &Not yet    \\ 
RFWash~\cite{khamis2020rfwash}     & 85\%         & 10        & sample-wise   &Not yet      \\ 
Uwash(Ours) & 92.27\%      & 51+10+10        & sample-wise & Yes\\ \bottomrule
\end{tabular}
\label{tab:sota}
\end{table}

\begin{figure}[t]
    \centering
    \includegraphics[width=1\linewidth]{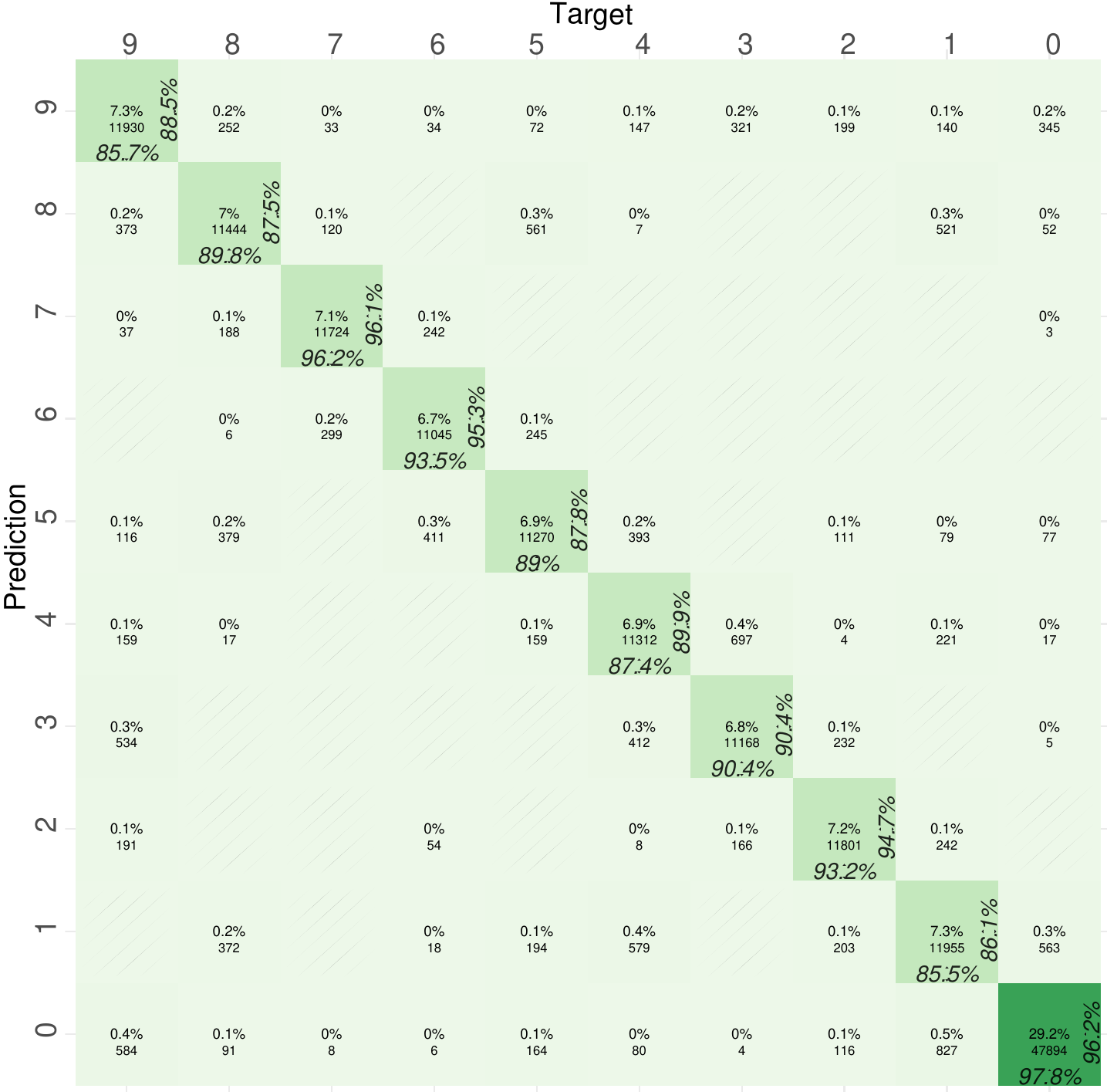}
    \caption{Confusion Matrix of Gesture Prediction. UWash works well for all gestures, especially for the $2$nd, $3$rd, $6$th, and $7$th gestures.}
    \label{fig:confusion}
\end{figure}

\begin{table}[t]
\centering
\caption{Error in Start/End Detection and Handwashing Scoring. $\downarrow$ means the less the better.}
\label{tab:startend}
\begin{tabular}{cccc}
\toprule
      & Start Detection & End Detection & Handwashing Scoring \\ \midrule
Mean $\downarrow$ & 0.49\textit{s}    & 0.23\textit{s} &  4.0pts \\
SD $\downarrow$  & 0.58\textit{s}    & 0.30\textit{s}  & 4.3pts \\ \bottomrule
\end{tabular}
\end{table} 

Next, we are going to show the performance of UWash from other four perspectives, i.e.,  performance on gestures, performance on participants, start/end detection error, and handwashing scoring results.

\textbf{(2) Performance on Gestures.} We show the confusion matrix of UWash on 10 gestures~(9 handwashing gestures + 1 background)~in Fig.~\ref{fig:confusion}. Though the data of these 10 gestures are not quite balanced~(29.2\% of background), UWash works well for all gestures, especially for the $2$nd, $3$rd, $6$th, and $7$th gestures. Besides, we find two types of false recognition are dominant. The first happens between the background and gestures of 1 and 9. We think this is because the background contains diverse activities such as wetting hands with water, applying soap, and drying hands with a towel, which may largely increase the difficulty to be classified correctly with its post-activity~(G1) or pre-activity~(G9). The other false recognition happens between two successive gestures. As the start/end time of every gesture is annotated manually, the discordance across different labeling workers, participants, and locations may cause false between successive gestures.

\begin{figure}[t]
    \centering
    \includegraphics[width=1\linewidth]{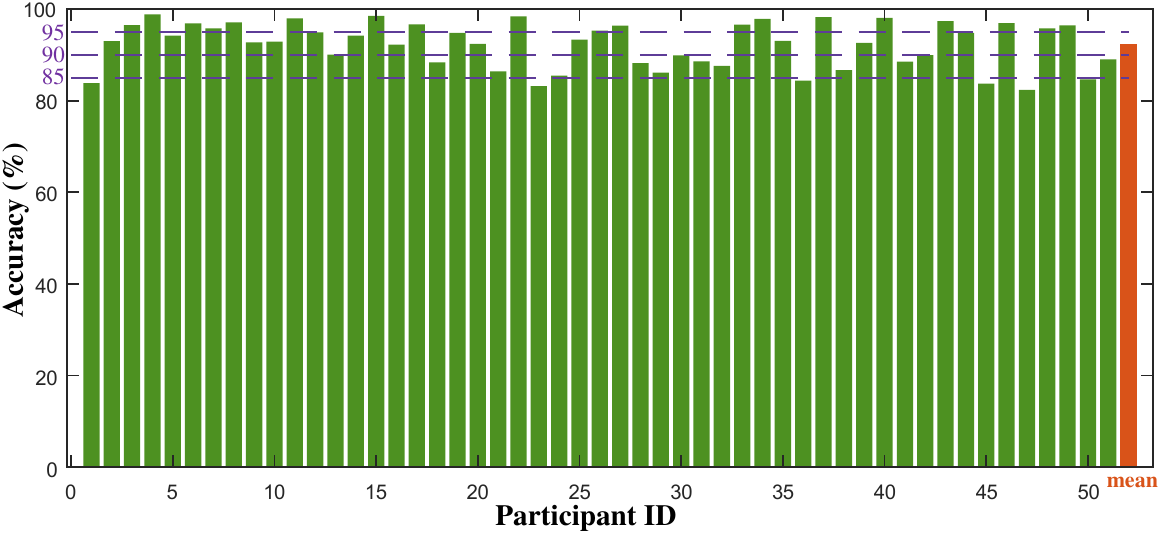}
    \caption{Results on participants. 33 out of 51 participants have an accuracy greater than 95\%, while 6 out of 51 are less than 85\% }
    \label{fig:userbar}
\end{figure}

\textbf{(3) Performance on Participants.} Fig.~\ref{fig:userbar} demonstrates the participant-wise accuracy, which shows that 33 of 51 participants achieve an accuracy of over 95\%; only 6 of them have an accuracy of less than 85\%; the average accuracy is 92.33\%~(orange bar). The results demonstrate that, for the seen participants, UWash achieves sample-wise handwashing gesture recognition effectively.

\begin{figure*}[t]
    \centering
    \includegraphics[width=1\linewidth]{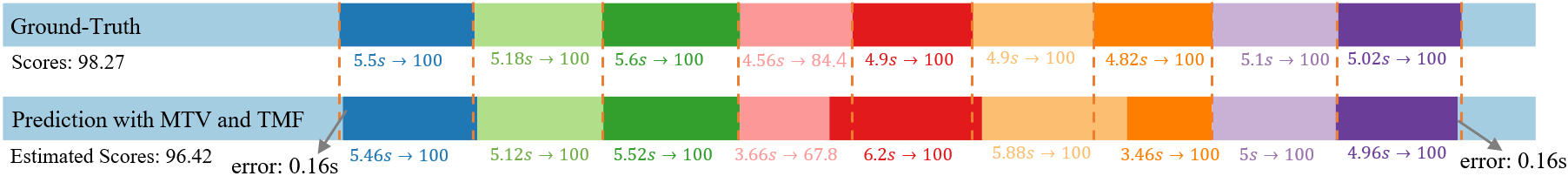}
    \caption{Visualization example of start/end detection and handwashing scoring.}
    \label{fig:visualization}
\end{figure*}

\textbf{(4) Start/End Detection.} Deserved to reiterate, UWash can automatically detect the handwashing start/end time, not requiring to work along with additional sensors~\cite{mondol2015harmony,samyoun2021iwash,zhong2016washindepth} or to be awakened manually~\cite{samyoun2021iwash}.  We compute the start/end detection errors over the test set and report the Mean and standard deviation~(SD) in Table~\ref{tab:startend}. The table shows that Means and SDs are all within 1 second, indicating UWash can detect handwashing events correctly and stably.

\textbf{(5) Scoring Results.} We report the handwashing scoring results in Table~\ref{tab:startend}. As shown in Table~\ref{tab:startend}, the mean and the SD are less than 5 points, indicating UWash can score handwashing well.

\textbf{(6) Visualization.}  In Fig.~\ref{fig:visualization}, we visualize how UWash works on a sample of the 26th participant. UWash outputs sample-wise gesture classification, with which we can detect the start/end time, estimate the duration of gestures, score gestures with the estimated duration, and score the whole handwashing procedure. Thus UWash can help users to promote their handwashing practice with the estimated scores in daily life.

\subsection{Cross-Domain Results}\label{sec:cu_results}
We evaluate the cross-domain performance of UWash, which is a critical criterion for a handwashing assessment system for people's daily use.

\begin{figure}[t]
    \centering
    \includegraphics[width=1\linewidth]{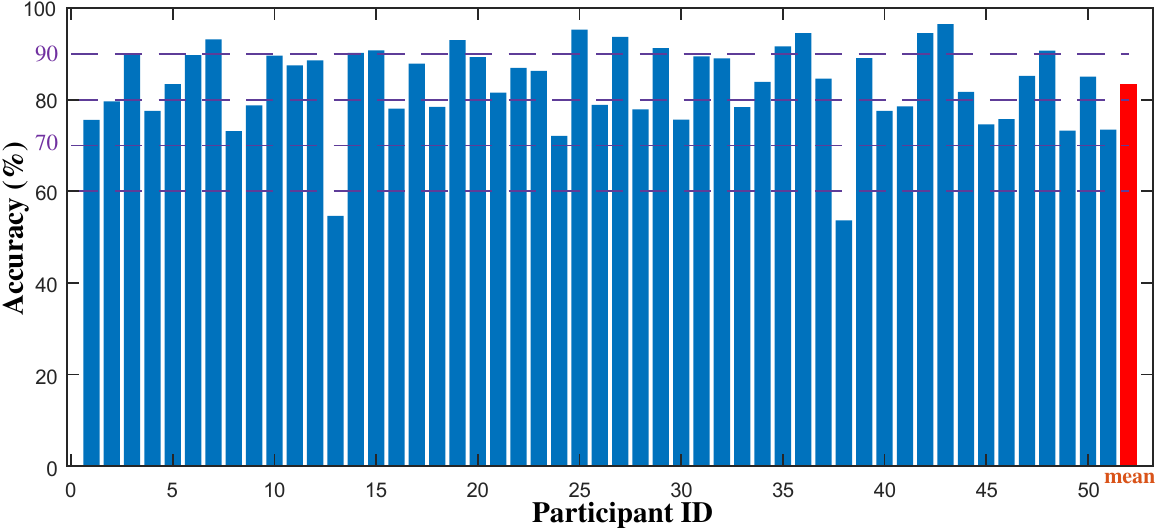}
    \caption{Cross-Participant Accuracy Results by Individual Participant: Accuracy for Different Participants}
 % Cross-Participant Results.
    \label{fig:crossuser}
\end{figure}

\begin{table}[t]
\centering
\caption{Cross-domain results(Average Accuracy and Errors). $\uparrow$ means the higher the better. $\downarrow$ means the less the better. CP, CP-CL, and CP-CL-CT are for  Cross-Participant, Cross-Participant-Cross-Location, and Cross-Participant-Cross-Location-Cross-Time.}
\begin{tabular}{lccc}
\toprule
         & CP & CP-CL &CP-CL-CT\\ \midrule
Accuracy$\uparrow$ & 83.34\%         & 81.45\% &82.08\%                            \\
Mean(start error)$\downarrow$     & 0.43\textit{s}         & 0.48\textit{s}                 &    0.39\textit{s}       \\
SD(start error)$\downarrow$     & 0.45\textit{s}         & 0.53\textit{s}              &    0.20\textit{s}              \\
Mean(end error)$\downarrow$       & 0.23\textit{s}         & 0.36\textit{s}            &    0.14\textit{s}                \\
SD(end error)$\downarrow$       & 0.33\textit{s}       & 0.44\textit{s}              &    0.09\textit{s}              \\
Mean(score error)$\downarrow$    & 12.74pts         & 15.21pts             &14.3pts               \\
SD(score error)$\downarrow$    & 8.01pts         & 9.26pts               &5.23pts            \\ \bottomrule
\end{tabular}
\label{tab:crossuser}
\end{table}

\textbf{(1) Cross-Participant.} We train UWash with data of 50 out of 51 participants and test the trained model with data of the remaining participant. We conduct this leave-one-participant-out process over 51 participants respectively to evaluate the cross-participant performance. As shown in Fig.~\ref{fig:crossuser}, accuracy among some left-out participants is more discrete than those in the user-dependent setting. For example, the 13th and the 38th participants have relatively special personal handwashing styles, resulting in an accuracy of $<$60\%. This indicates that they may not follow the WHO guidelines. In this case, UWash could remind users to improve their hand hygiene techniques. As expected, the mean accuracy in the cross-participant setting decreases to 83.34\%, since the personalized gestures of these individuals are not included in the training set. 

Table~\ref{tab:crossuser} shows that UWash detects the start/end time of handwashing well even in the cross-participant setting, with errors of $<$0.5~\textit{seconds}, which means that UWash can effectively distinguish the handwashing gestures and other activities. However, the handwashing scoring performance drops significantly, repeatedly indicating the performance of gesture classification on unseen users highly depends on how well they wash hands following WHO guidelines.

\textbf{(2) Cross-Participant-Cross-Location.} We use data from 4 out of 5 locations to train UWash and test the trained model on the remaining one location. We conduct this leave-one-location-out process over 5 locations where data is collected respectively. Since recruited participants have no overlap between different locations, the leave-one-location-out process also leads to the evaluation in the cross-participant-cross-location setting.

\begin{figure}[t]
    \centering
    \includegraphics[width=1\linewidth]{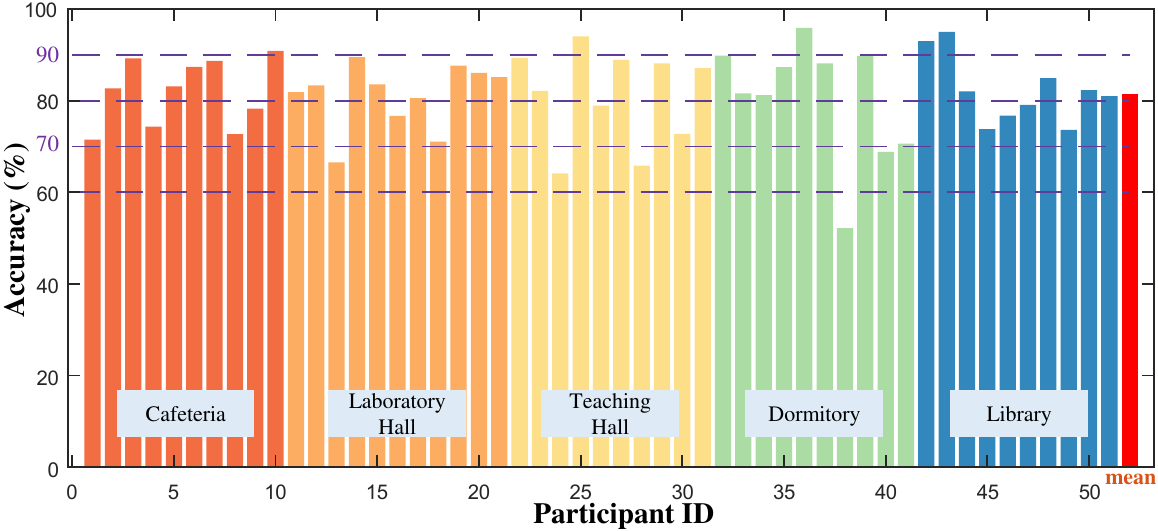}
    \caption{Cross-Participant-Cross-Location Accuracy by Individual Participant: Accuracy Distribution across Locations.}
    \label{fig:crosslocationbar}
\end{figure}

The experimental results are shown in Fig.~\ref{fig:crosslocationbar} and Table~\ref{tab:crossuser}. In this setting, the performances have similar characteristics to those in the cross-participant setting, e.g., the accuracy is more discrete; the 13th and the 38th participants have the lowest accuracy; the mean accuracy drops to 81.45\%; the start/end time detections are achieved well.

We further find that the performance in the cross-participant setting is slightly better than those in the cross-participant-cross-location setting. We think this is because we use data from 50 participants to train UWash each time in the former setting, while in the latter setting we use less data of 40 or 41 participants from four locations to train UWash each time.

\textbf{(3) Cross-Participant-Cross-Location-Cross-Time.} Nine months after the data collection described in Sec.~\ref{sec:data}, we further randomly recruit 10 passersby in a hospital to wash their hands 5 times, and follow the same process to obtain a new dataset. We apply the trained model in user-dependent results on the new dataset to evaluate the cross-time performance. 

\begin{table*}[t]
\centering
\caption{Cross-Participant-Cross-Location-Cross-Time Results for Individual Participants. UWash is stable and promising across time.}
\label{tab:testoftime}
\begin{tabular}{lccccccccccc}
\toprule
Passersby ID   & 1 & 2 & 3 & 4 & 5 & 6 & 7 & 8 & 9 & 10 & \textbf{Avg.} \\ \midrule
Accuracy~(\%)$\uparrow$ & 89.02 & 87.16 & 78.20 & 72.47 & 93.38 & 95.12 & 57.23 & 92.03 & 69.86 & 86.31 & \textbf{82.08} \\
Mean(start error, \textit{s})$\downarrow$   & 0.10 & 0.33 & 0.54 & 0.58 & 0.21 & 0.08 & 1.26 & 0.13 & 0.39 & 0.24 & \textbf{0.39} \\
SD(start error, \textit{s})$\downarrow$     & 0.07 & 0.22 & 0.14 & 0.33 & 0.11 & 0.08 & 0.65 & 0.11 & 0.16 & 0.13 & \textbf{0.20} \\
Mean(end error, \textit{s})$\downarrow$     & 0.16 & 0.14 & 0.06 & 0.22 & 0.26 & 0.08 & 0.20 & 0.12 & 0.10 & 0.04 & \textbf{0.14} \\
SD(end error, \textit{s})$\downarrow$       & 0.10 & 0.04 & 0.06 & 0.23 & 0.10 & 0.04 & 0.14 & 0.11 & 0.09 & 0.03 & \textbf{0.09} \\
Mean(score error, pts)$\downarrow$   & 10.95 & 11.48 & 16.78 & 23.85 & 5.19 & 3.83 & 36.83 & 5.71 & 21.56 & 6.86 & \textbf{14.30} \\
SD(score error, pts)$\downarrow$     & 1.77 & 8.83 & 3.70 & 10.66 & 1.64 & 3.21 & 4.45 & 5.27 & 8.59 & 4.14 & \textbf{5.23} \\ \bottomrule
\end{tabular}
\end{table*}

Table~\ref{tab:testoftime} shows the cross-time results of the 10 passersby. The average sample-wise gesture recognition accuracy is 82.08\%. The average error in the start and end detection of handwashing are 0.39\textit{s} and 0.14\textit{s} respectively. The average standard deviation in the start and end detection of handwashing are 0.20\textit{s} and 0.09\textit{s} respectively. The average scoring error and scoring standard deviation are 14.30 points and 5.23 points respectively. We find that these average values are at the same level as in the Cross-Participant-Cross-Location evaluation listed in Table~\ref{tab:crossuser}. These results demonstrate that UWash is stable and promising across time. Note that the new dataset is from the new participants, new location, and new date, results of this evaluation provide us strong confidence in performance if UWash is promoted to large-scale real-world use.

\begin{figure}[t]
    \centering
    \includegraphics[width=1\linewidth]{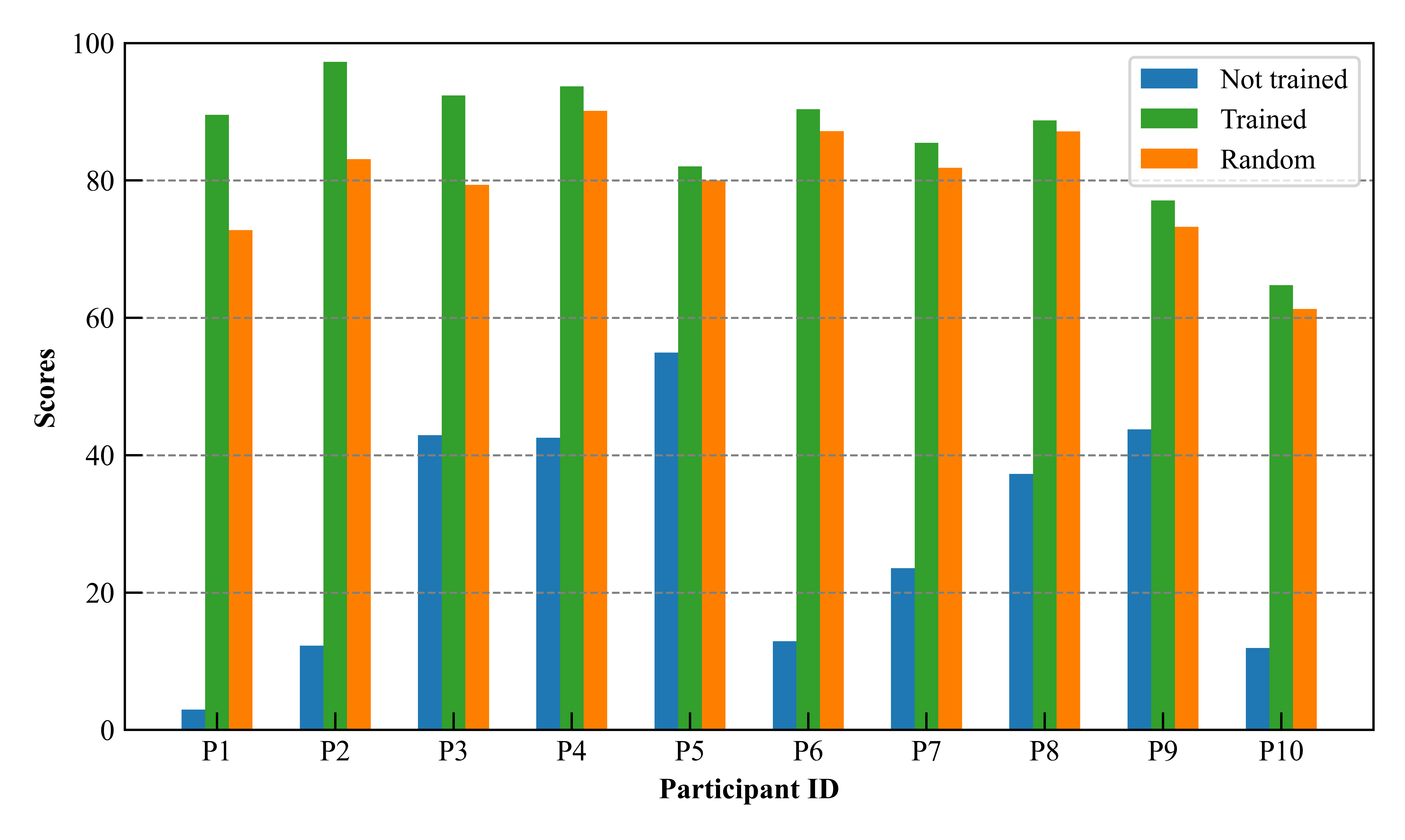}
    \caption{The Wild Study. UWash can work under the conditions when people shuffle the handwashing steps or omit steps.}
    \label{fig:wildtest}
\end{figure}

\textbf{(4) The Wild Study.} To test UWash in the wild, we further recruit 10 passersby in a laboratory hall to evaluate their handwashing quality under three conditions: their usual handwashing method without prior training on the WHO-recommended technique (not trained), after receiving our training on the WHO handwashing guideline (trained), and when they intentionally shuffle the order of the WHO handwashing steps or omit certain steps (random). Five trials are collected for each individual and condition. We directly apply the UWash model trained in the in-domain evaluation section (Sec.~\ref{sec:results}) to estimate the handwashing scores,  and the average scores are presented in Fig.~\ref{fig:wildtest}.

As shown in Fig.~\ref{fig:wildtest}, almost all not-trained participants have low handwashing scores, indicating that they do not adhere to the WHO-recommended handwashing procedure in their daily life. After training, their handwashing scores significantly improve, reaching an average score of approximately 85. When participants deliberately shuffle the handwashing steps or omit steps, UWash can detect a decline in their scores, demonstrating its effectiveness in identifying out-of-sequence gestures or missed steps. In practical use, UWash can alert users to any omitted steps, thereby enhancing handwashing quality in a targeted manner.

It is worth noting that participants P4 to P10 only shuffle the order of handwashing steps without omitting any, and their scores should have been close to those obtained during the trained session. However, these scores were slightly lower because, during the random session, they did not strictly adhere to the prescribed duration for handwashing. Since UWash evaluates scores based on the duration spent on each step, this deviation from the recommended durations leads to a reduction in overall scores.

\section{Discussion and Limitation}~\label{sec:discussion}

There are still several aspects of UWash that need further discussion.

\textbf{(1) Handwashing Correctness:}  The current UWash system reduces hand hygiene quality assessment to merely the duration of handwashing actions, neglecting critical factors such as thorough coverage of surfaces, correct motions, soap coverage, and scrubbing vigor that significantly impact cleaning efficacy. A potential solution involves leveraging parametric hand pose or mesh estimation techniques to characterize fine-grained hand movements. For instance, employing the MANO model~\cite{romero2022embodied} for hand pose estimation allows for more precise modeling of hand postures. Standalone smartwatches may struggle to capture comprehensive hand pose. A recent work, Ring-a-Pose~\cite{yu2024ring}, demonstrates high-accuracy hand pose estimation through a smart ring worn on the middle finger. Future iterations of UWash could collaborate with devices like smart rings to acquire hand pose or mesh, thereby enabling more precise hand hygiene quality assessments.

\textbf{(2) Smartwatch Placement: }  The current UWash evaluation assumes a standard watch orientation during testing. However, real-world usage introduces variability—such as loose-fitting wear, inner-wrist positioning, or placement on the non-dominant arm—that challenges system robustness. To address this, additional data collection under such conditions could be implemented, though this approach demands significant human labor. A more labor-efficient alternative involves leveraging synthetic data by establishing a hand coordinate system and injecting orientation-specific perturbations into IMU signals to replicate coordinate system instabilities across distinct wearing positions and varying tightness levels. Domain adaptation~\cite{ganin2016domain} and domain generalization~\cite{li2018learning} techniques could also be explored to mitigate these challenges.

\begin{table*}[t]
\centering
\caption{UWash latency on a 2010-launched Intel Core i7-620M desktop (seconds). }
\label{tab:cpu}
\begin{tabular}{cccccc}
\toprule
 stride  & \textbf{clip inference  time} &  sequence inference time & post-processing time &  score calculation time & \textbf{total time }    \\ \midrule
1  & 0.0067±0.0000 & 20.4945±4.0483 & 0.3477±0.0680 & 0.0026±0.0004 & 20.8448±4.1167 \\
10 & 0.0067±0.0002 & 2.0488±0.3920  & 0.1319±0.0256 & 0.0025±0.0004 & 2.1832±0.4177  \\
20 & 0.0067±0.0003 & 1.0311±0.2051  & 0.1207±0.0234 & 0.0025±0.0004 & 1.1543±0.2284  \\
30 & 0.0067±0.0000 & 0.6789±0.1321  & 0.1169±0.0225 & 0.0026±0.0004 & 0.7984±0.1550  \\
40 & 0.0068±0.0003 & 0.5185±0.0983  & 0.1141±0.0222 & 0.0025±0.0004 & 0.6351±0.1205  \\
50 & 0.0067±0.0002 & 0.4131±0.0811  & 0.1139±0.0223 & 0.0025±0.0004 & 0.5295±0.1035  \\
64 & 0.0067±0.0000 & 0.3207±0.0619  & 0.1113±0.0217 & 0.0025±0.0004 & 0.4345±0.0840 \\ \bottomrule
\end{tabular}
\end{table*}

\textbf{(3) Power Consumption and Latency:} UWash has not been deployed on smartwatches for real-world evaluation of battery consumption and inference latency, despite its parameter count being only 3.2\% of MobileNetV3-small~\cite{koonce2021mobilenetv3}, a classic deep network optimized for edge devices. In practical deployment, the UWash system architecture can be designed as follows: On the smartwatch end, IMU sensor data is sampled and transmitted to a paired smartphone via Bluetooth Low Energy (BLE). The IMU module, operating at 5 mW (typical for MEMS sensors in continuous sampling mode), and the BLE transmission, consuming 8 mW under a 1 Mbps data rate with 1 ms connection intervals, combine for a total power draw of 13 mW, enabling a smartwatch with a 300 mAh battery capacity (e.g., Samsung Gear Sport) to achieve approximately 115.4 hours (5V × 0.3Ah/0.013W $\approx$ 115.4h) of continuous operation. Meanwhile, the smartphone handles real-time handwashing assessment using the PyTorch Mobile inference engine and relays the results back to the smartwatch via BLE for user feedback. 

We evaluated UWash on a desktop-class CPU, the i7-629M (released in 2010), simulating 50 iterations of 60-second handwashing sequences. We measured the average network inference time, post-processing time, and computational steps for scoring, with results summarized in Table~\ref{tab:cpu}. As shown in Table~\ref{tab:cpu}, with a sliding window stride of one sample point during inference, the total processing time was 20.8448±4.1167 seconds, dominated by the sequence inference time of 20.4945±4.0483 seconds. When the stride was increased to 10, the total time decreased to 2.1832±0.4177 seconds, demonstrating that UWash can process a 60-second sequence in approximately 2.1832±0.4177 seconds. For clip-level inference, UWash processes a 1.28-second IMU clip in just 0.0067 seconds.

According to benchmarking data from \url{https://www.cpubenchmark.net/cpu_list.php}, the Intel Core i7-629M performs comparably to mobile chips like the Qualcomm Snapdragon 835 and Apple A9. Representative devices using these chips include the 9-year-old Samsung Galaxy S8 and Xiaomi 6 (Snapdragon 835), and the 11-year-old Apple iPhone 6S (A9). These results give a rough estimate of feasibility that mainstream smartphones are capable of supporting real-time execution of UWash, but actual deployment performance may vary due to real-world conditions.

\textbf{(4) Other Brands of Smartwatches: } To evaluate the UWash on other brands of smartwatches, we further recruited 10 new subjects and instructed them to perform five handwashing sessions while wearing the Apple Watch S5. Due to discrepancies in the IMU sensor coordinate systems between the Apple Watch S5 and Samsung Gear Sport, combined with the lack of publicly available parameters for their respective coordinate frameworks, direct parameter alignment remains a significant challenge. Nevertheless, through simple axis swapping—specifically interchanging the x- and y-axes in the Apple Watch data—we tested the Samsung-trained model on Apple Watch data, achieving cross-device performance metrics of 76.24\% accuracy, 0.39s start error, 0.57s end error, and 18.81-point scoring error. These results highlight the generalizability of the UWash framework across heterogeneous devices. To improve alignment accuracy, future implementations would require obtaining precise transformation matrices, including the rotation matrix and translation parameters.

Different smartwatch models may exhibit varying IMU sampling rates. For devices with sampling rates below 50 Hz (the default sampling rate of UWash), upsampling techniques such as linear interpolation could be applied to achieve the target 50 Hz frequency prior to model inference. However, this compensatory strategy inherently fails to recover critical kinematic details lost during rapid hand motion transitions, which clinically manifests as reduced detection fidelity in hand hygiene monitoring. We also could downsample the training data to match the target device’s native sampling rate during model training. For devices with sampling rates exceeding 50 Hz, raw data can be downsampled to 50 Hz to align with UWash.

\textbf{(5) Border Impacts:} Our proposed framework is a general-purpose foundation for action recognition and segmentation, with applicability extending far beyond the domain of handwashing. By integrating agent technologies and advanced reasoning capabilities, the system can evolve beyond passive monitoring to deliver actionable insights. For instance, it could generate behavioral summaries to track long-term skill development, provide context-aware feedback tailored to individual user patterns~\cite{lan2025xrf}, or offer real-time corrective guidance during task execution~\cite{arakawa2024prism}.  This integration not only improves the system's utility but also opens avenues for personalized skill augmentation, adaptive training interfaces, and human-AI collaboration in complex workflows.

\section{Conclusion}~\label{sec:conclusion}

In this paper, we introduce UWash, a smartwatch-only handwashing assessment system, to raise people's awareness of handwashing in daily use and adherence to the WHO handwashing guidelines. UWash takes the data of the IMU sensors of smartwatches, i.e., accelerometers and gyroscopes, as inputs, feeds the inputs into dual-branch U-Nets, and outputs sample-wise gesture recognition results effectively.  UWash can detect handwashing start/end time, estimate the duration of every handwashing gesture, and score gestures as well as the entire procedure following WHO guidelines. Experimental results over 51 participants show that UWash works well in the user-dependent, cross-participant, and cross-participant-cross-location settings. In addition, UWash still performs promising 9 months later in a hospital and 2 years later in a hall during an in-wild test. Moreover, UWash models are lightweight, only 496KB, with great potential to be deployed on edge devices in the future.

\bibliographystyle{IEEEtran}
\bibliography{./reference}

\vspace{-35pt}
\begin{IEEEbiography}[{\includegraphics[width=1in,height=1.25in,clip,keepaspectratio]{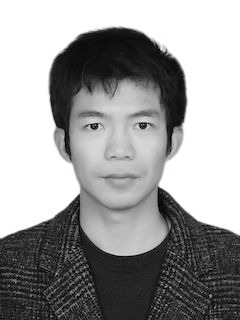}}]
{Fei Wang} (Member, IEEE) received the B.E. and Ph.D. degree in Computer Science and Technology from Xi’an Jiaotong University, Xi'an, China, in 2013 and 2020, respectively. He was a visiting Ph.D. student with the School of Computer Science, Carnegie Mellon University, Pittsburgh, USA, from 2017 to 2019. He is currently an Associate Professor with Xi’an Jiaotong University. His research interests include human sensing, mobile computing, and deep learning.

\end{IEEEbiography}

\vspace{-35pt}
\begin{IEEEbiography}[{\includegraphics[width=1in,height=1.25in,clip,keepaspectratio]{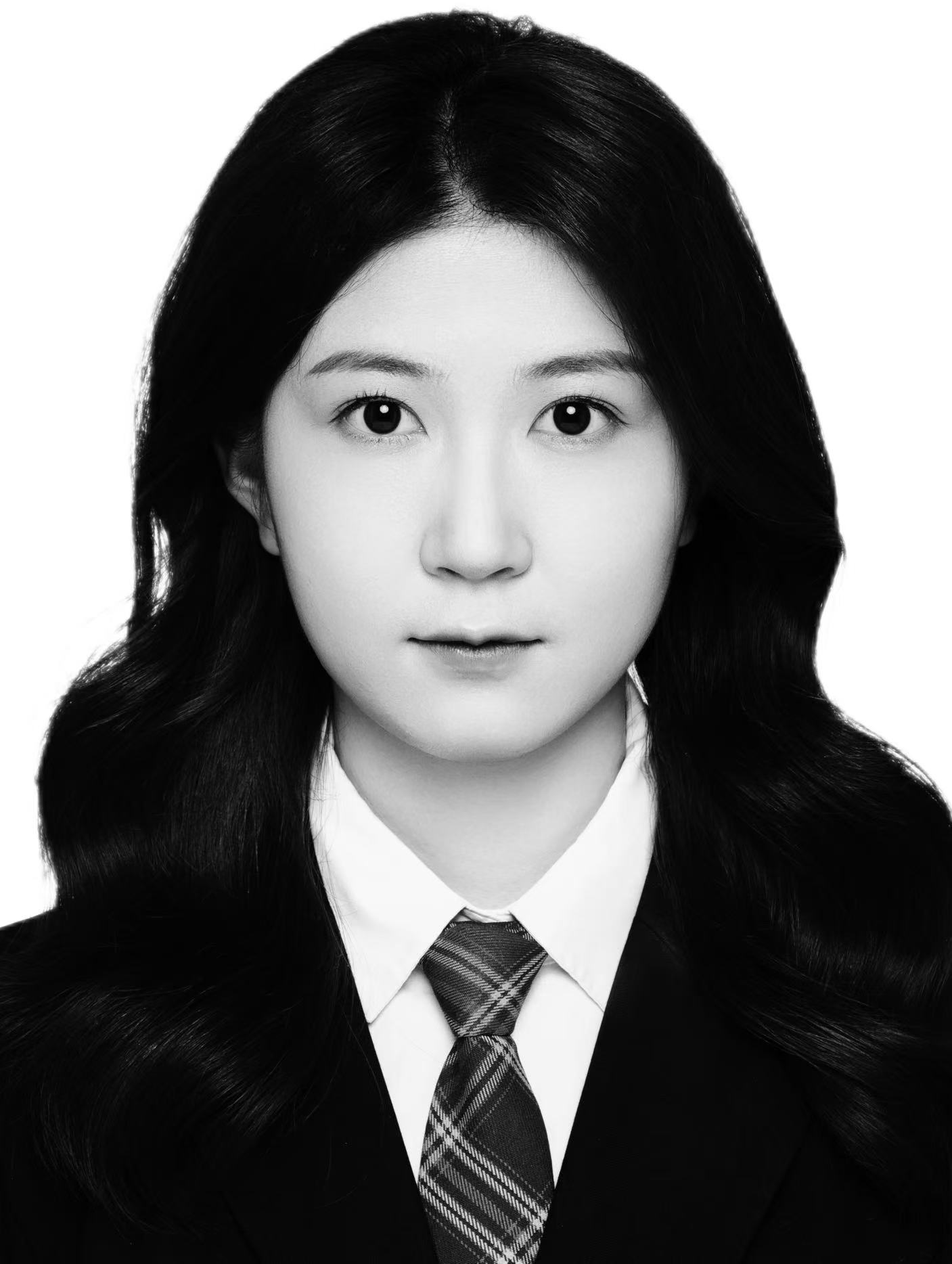}}]
{Tingting Zhang} received the B.E. degree from the School of Information Technology, Xiamen University, Xiamen, China, in 2024. She is currently a master student with the School of Software Engineering, Xi'an Jiaotong University, Xi'an, China. Her research interest is human-computer interaction.
\end{IEEEbiography}

\vspace{-35pt}
\begin{IEEEbiography}[{\includegraphics[width=1in,height=1.25in,clip,keepaspectratio]{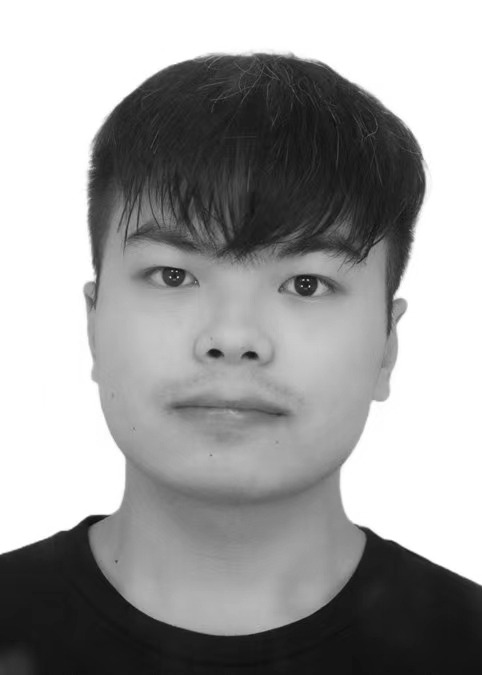}}]
{Xilei Wu} received the M.E. degree from the School of Software Engineering, Xi'an Jiaotong University, Xi'an China, in 2023. He is currently a Software Engineer with Kuaishou Technology Inc., Beijing, China.
\end{IEEEbiography}

\vspace{-35pt}
\begin{IEEEbiography}[{\includegraphics[width=1in,height=1.25in,clip,keepaspectratio]{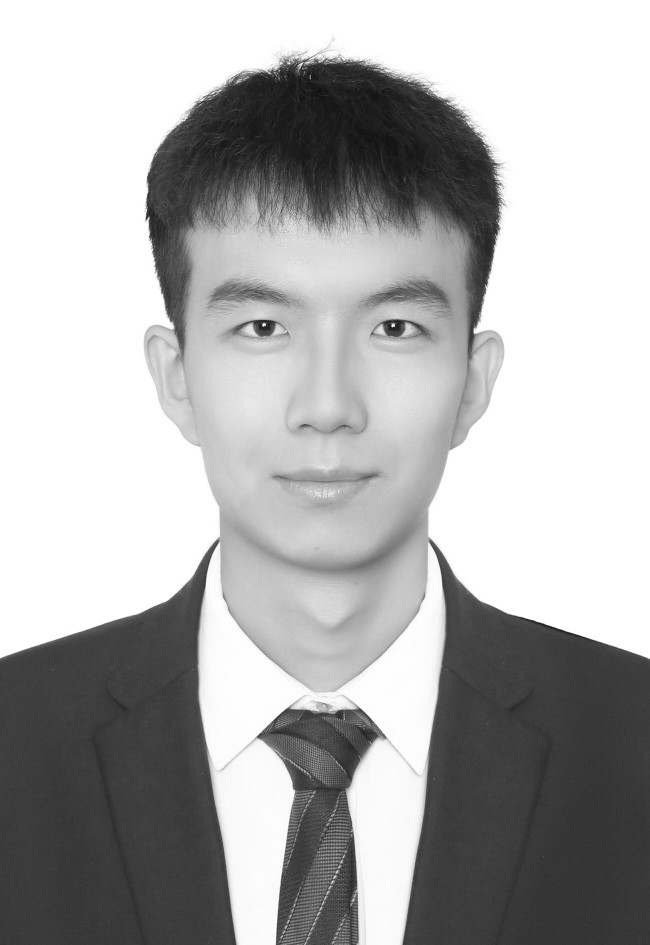}}]
{Pengcheng Wang}  received the B.S. degree from the School of Aeronautics and Astronautics, Shenyang Aerospace University, Shenyang China, in 2022. He is currently a master student with the School of Software Engineering, Xi'an Jiaotong University, Xi'an, China. His research interest is mobile computing.
\end{IEEEbiography}

\vspace{-35pt}

\begin{IEEEbiography}[{\includegraphics[width=1in,height=1.25in,clip,keepaspectratio]{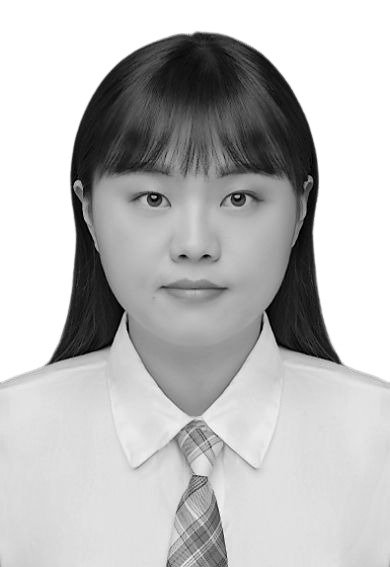}}]
{Xin Wang} received the M.E. degree from the School of Software Engineering, Xi'an Jiaotong University, Xi'an, China, in 2023.
\end{IEEEbiography}

\vspace{-35pt}

\begin{IEEEbiography}[{\includegraphics[width=1in,height=1.25in,clip,keepaspectratio]{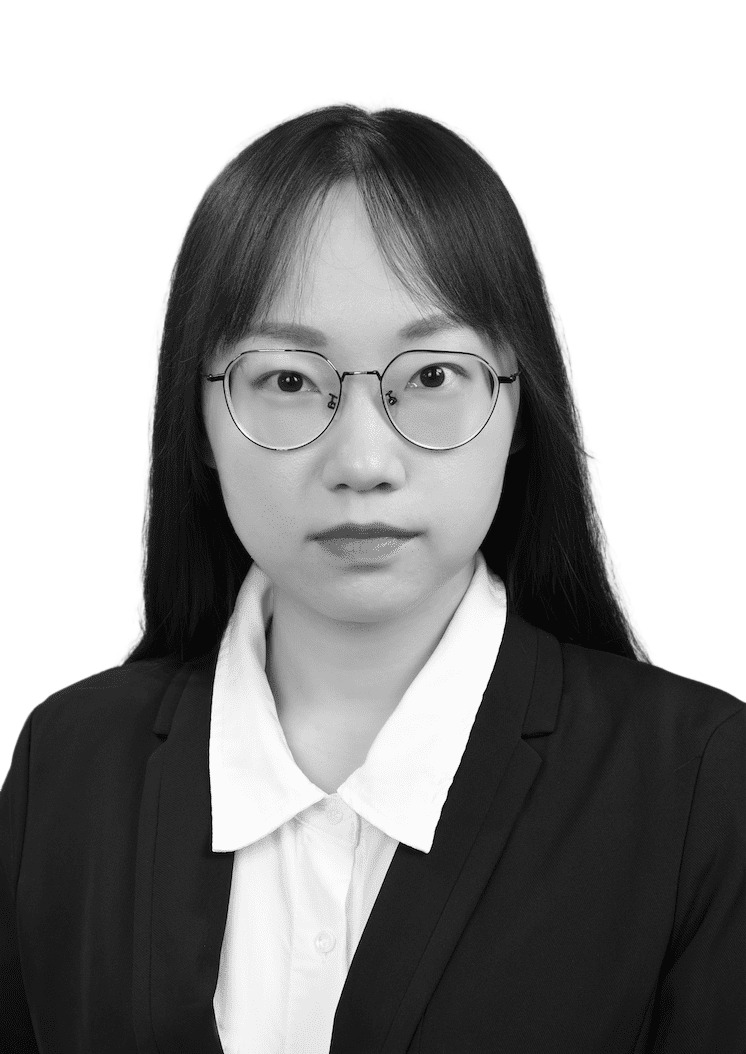}}]
{Han Ding} (Senior Member, IEEE) received the Ph.D. degree in computer science and technology from Xi’an Jiaotong University, Xi’an, China, in 2017.
She is currently an Associate Professor with Xi’an Jiaotong University. His research interests focus on AIoT, smart sensing, and RFID systems.
\end{IEEEbiography}
\vspace{-35pt}

\begin{IEEEbiography}[{\includegraphics[width=1in,height=1.25in,clip,keepaspectratio]{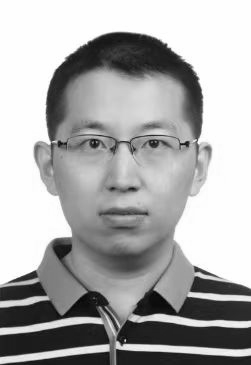}}]
{Jingang Shi} received the B.S. and Ph.D. degrees from the Department of Electronics and Information Engineering, Xi’an Jiaotong University, China. From 2017 to 2020, he was a Postdoctoral Researcher with the Center for Machine Vision and Signal Analysis, University of Oulu, Finland. Since 2020, he has been an Associate Professor with the School of Software, Xi’an Jiaotong University. His current research interests mainly include biomedical signal processing.
\end{IEEEbiography}

\vspace{-35pt}

\begin{IEEEbiography}[{\includegraphics[width=1in,height=1.25in,clip,keepaspectratio]{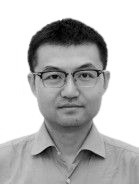}}]
{Jinsong Han} (Senior Member, IEEE) received his Ph.D. degree from Hong Kong University of Science and Technology in 2007. He is now a professor with Zhejiang University, Hangzhou, China. He is a senior member of the ACM and IEEE. His research interests focus on IoT security, smart sensing, wireless and mobile computing.
\end{IEEEbiography}

\vspace{-35pt}
\begin{IEEEbiography}[{\includegraphics[width=1in,height=1.25in,clip,keepaspectratio]{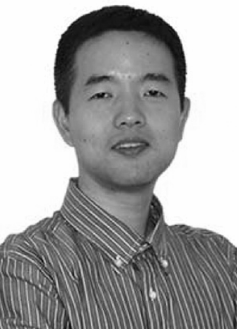}}]
{Dong Huang} is a senior project scientist at the Robotics Institute at Carnegie Mellon University, USA. Since 2018, he directs the DeLight Lab at Carnegie Mellon
University ( \url{https://www.ri.cmu.edu/robotics-groups/delight/}). His research focuses on deep learning perception on multi-modality systems and embedded platforms. He received his M.Sc. in Automation and PhD degrees in Computer Science from University of Electronic Science and Technology of China, respectively, in 2005 and 2009, Chengdu, China. 
\end{IEEEbiography}

\clearpage

\appendices
% \nopagenumbers
% \IEEEPARstart{The Online Questionnaire}

\section{The Online Questionnaire}\label{sec:questionnaire}

\begin{figure*}[b]
    \centering
    \includegraphics[width=1\linewidth]{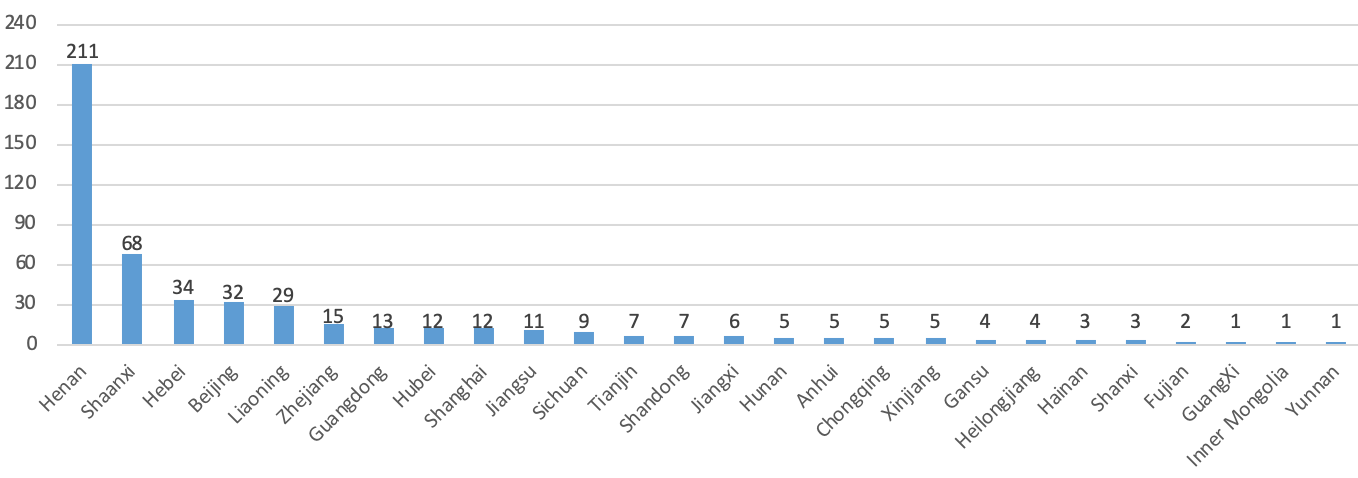}
    \caption{We conducted an online questionnaire on handwashing knowledge and practices over 505 subjects across 26 provinces in China.}
    \label{fig:provinces}
\end{figure*}

\begin{figure*}[b]
    \centering
    \includegraphics[width=1\linewidth]{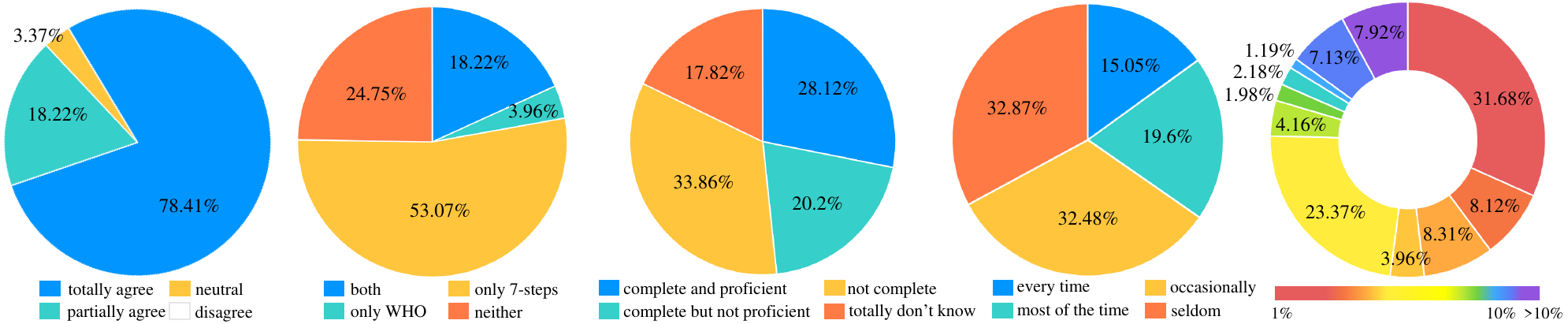}
    \caption{Online questionnaire results over 505 subjects, corresponding to Q1 to Q5 respectively.}
    \label{fig:userstudy}
\end{figure*}

In the 1st paragraph of Sec.Introduction, we mentioned that we had conducted an online questionnaire on handwashing knowledge and practices over 505 subjects across 26 provinces in China. Fig.~\ref{fig:provinces} lists the number of subjects from these provinces.  We report the results of five questions in the questionnaire in Fig.~\ref{fig:userstudy}, leaving out information on the age, gender, job, etc. The first four pie charts correspond to the questions below respectively.

\textit{ Q1. Handwashing is very important in daily life.}

\textit{ Q2. Have you heard of the WHO handwashing guidelines or the 7-step guidelines?}

\textit{ Q3. How well do you master each step of the WHO handwashing guidelines or the 7-step guidelines?}

\textit{ Q4. How often do you follow the WHO handwashing guidelines or the 7-step guidelines when washing hands?}

These charts show that 96.63\% of subjects think handwashing is important in daily life. Meanwhile, 96.04\% of subjects have heard of standard handwashing guidelines. However, 48.32\% of subjects know all steps. What's worse, only 34.65\% subjects wash hands following standard guidelines most of the time. The statistics show that there is a big gap between the knowledge and practices on handwashing in people's daily life, indicating the urgency to propose an automatic handwashing monitoring system for daily use.

\textit{Q5. Relative to the price of a smartwatch, how much are you willing to pay for an App of handwashing scoring?}

The target of Q5 is to investigate the purchase will of potential users. The last chart in Fig.~\ref{fig:userstudy} shows that 31.68\% of subjects will pay an additional 1\% price for the handwashing monitoring services~(\$3.99 for Apple Watch S6 and Samsung Galaxy Watch3); and 23.37\% of subjects will pay an additional 5\%. Surprisingly, 15.05\% of subjects are willing to pay an additional 10\% or more. These statistics show that handwashing monitoring apps like UWash have a considerable market value.

% \section{Hand-washing Steps Videos at Youtube}\label{sec:video_urls}
%  we carefully selected 12 videos and recorded the duration of each handwashing stage. We then averaged these durations to establish a gold standard for stage durations, which serves as the basis for our
% scoring.

% \noindent (1) \url{https://www.youtube.com/watch?v=qo7Q_wm2Vec}\\
% (2) \url{https://www.youtube.com/watch?v=IisgnbMfKvI}\\
% (3) \url{https://www.youtube.com/watch?v=0FLQ-EpQ6PM}\\
% (4) \url{https://www.youtube.com/watch?v=jXqDAfeUFBg}\\
% (5) \url{https://www.youtube.com/watch?v=6JrEeR5OXiE}\\
% (6) \url{https://www.youtube.com/watch?v=TClRYmtqClM}\\
% (7) \url{https://www.youtube.com/watch?v=a9CMtzymZTg}\\
% (8) \url{https://www.youtube.com/watch?v=3PmVJQUCm4E}\\
% (9) \url{https://www.youtube.com/watch?v=hhKlyoVsbOY}\\
% (10) \url{https://www.youtube.com/watch?v=3KjUaibd4gg}\\
% (11) \url{https://www.youtube.com/watch?v=4CcGLoYrIPU}\\
% (12) \url{https://www.youtube.com/watch?v=YiChdJ_os3Q&t=53s}

\clearpage
\section{Handwashing Assessment Examples}\label{sec:results_show}

\begin{figure*}[b]
    \centering
    \includegraphics[width=0.9\linewidth]{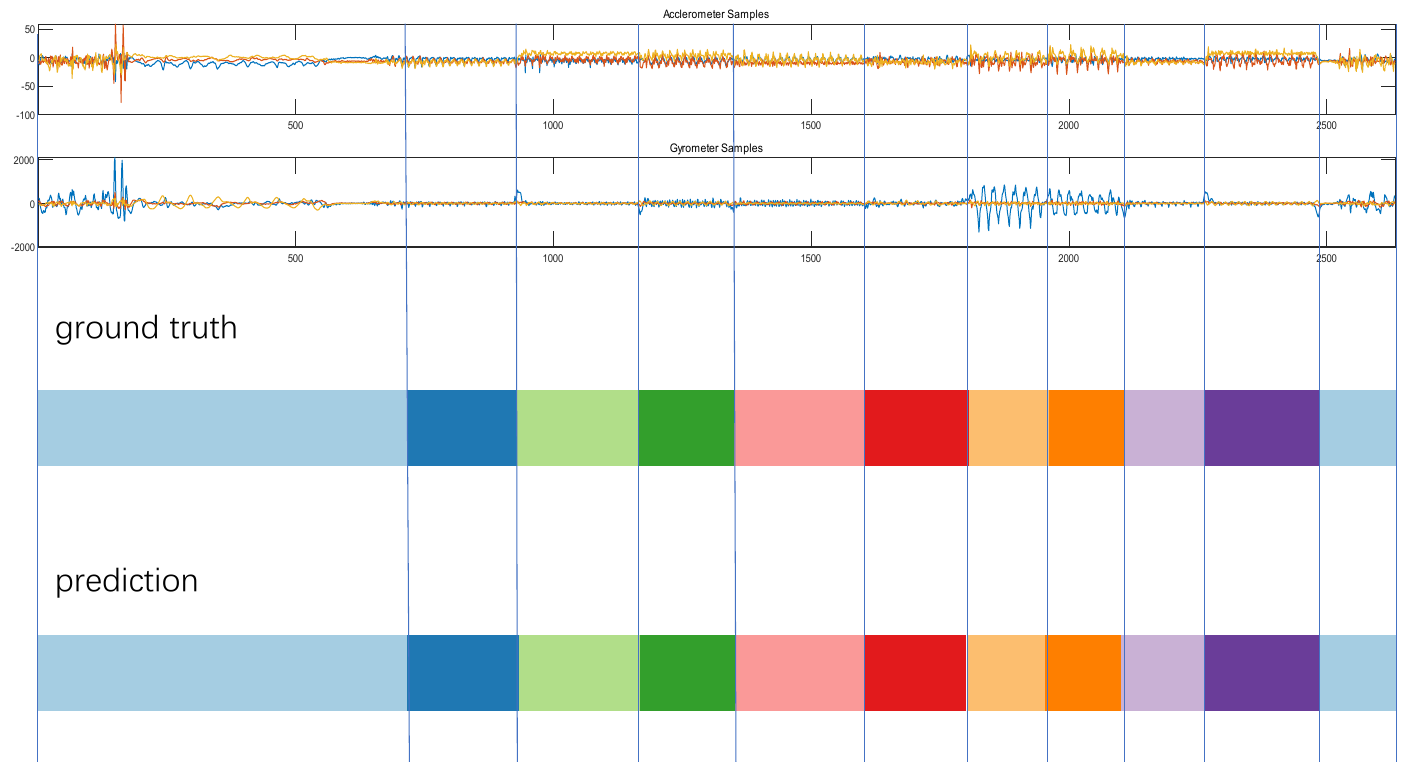}
    \caption{Results of the 7th subject in the cafeteria.}
    \label{fig:canteen7}
\end{figure*}

\begin{figure*}[b]
    \centering
    \includegraphics[width=0.9\linewidth]{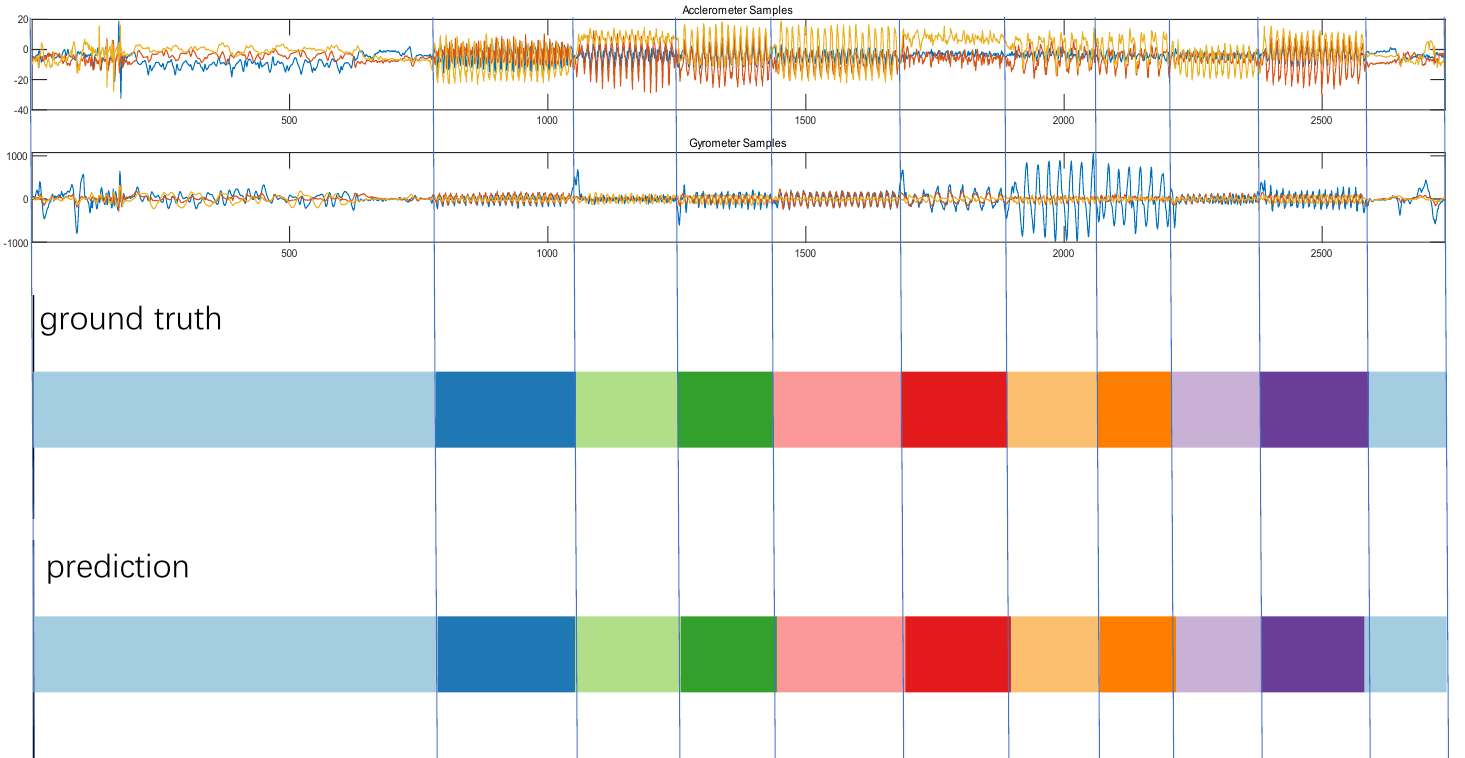}
    \caption{Results of the 10th subject in the cafeteria.}
    \label{fig:canteen_10}
\end{figure*}

\begin{figure*}[t]
    \centering
    \includegraphics[width=1\linewidth]{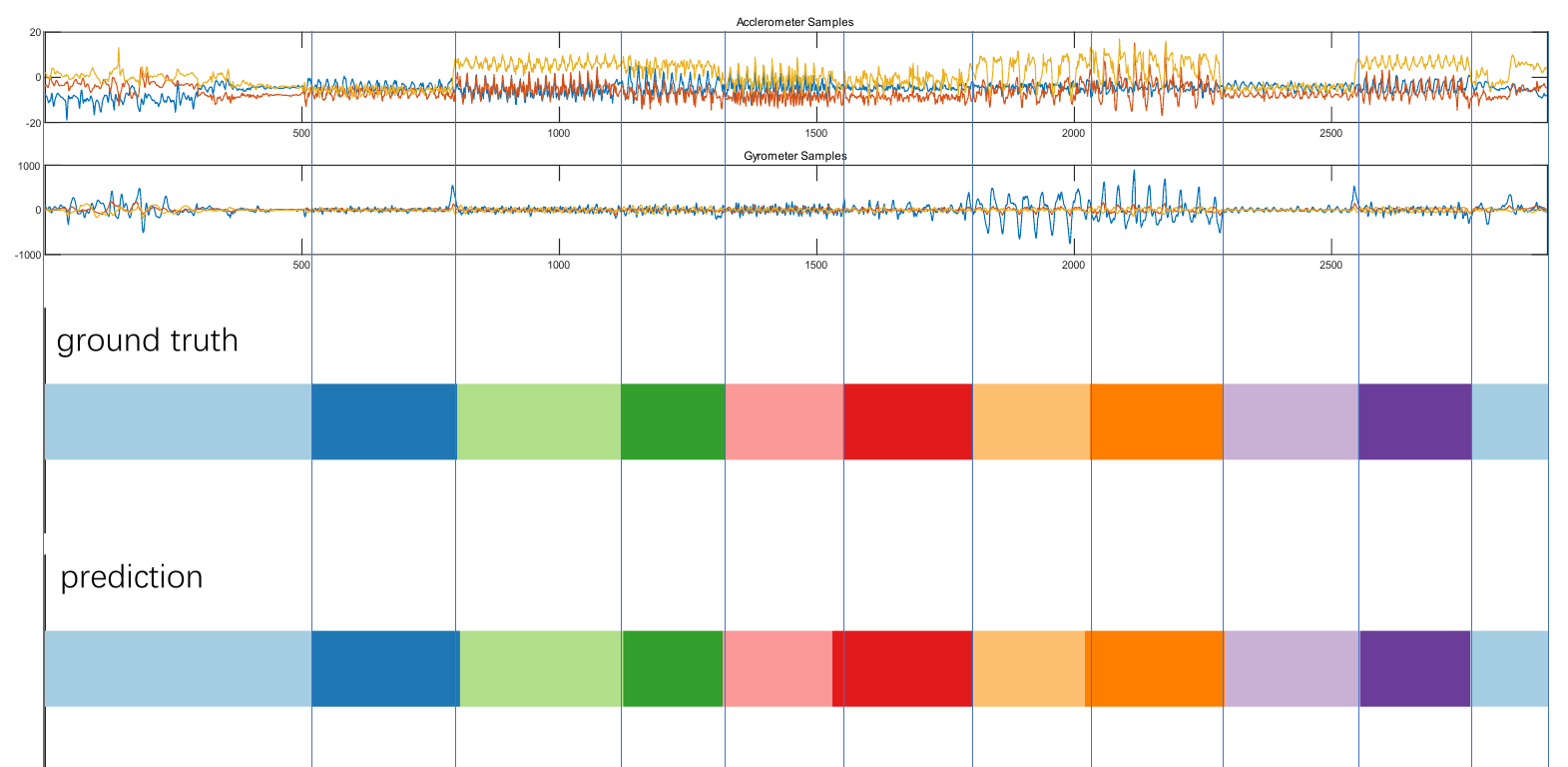}
    \caption{Results of the 4th subject in the dormitory.}
    \label{fig:dormitory_4}
\end{figure*}

\begin{figure*}[t]
    \centering
    \includegraphics[width=1\linewidth]{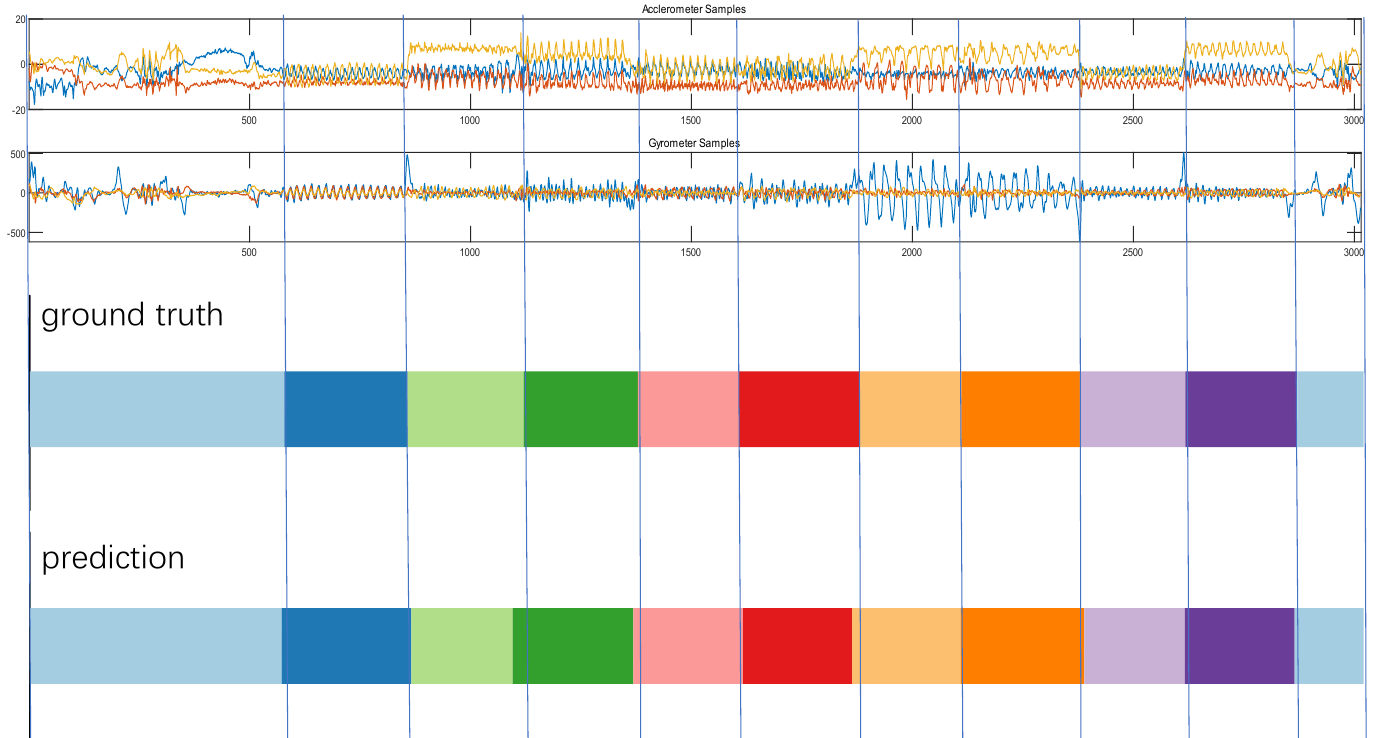}
    \caption{Results of the 8th subject in the teaching hall.}
    \label{fig:hanying_8}
\end{figure*}

\begin{figure*}[t]
    \centering
    \includegraphics[width=1\linewidth]{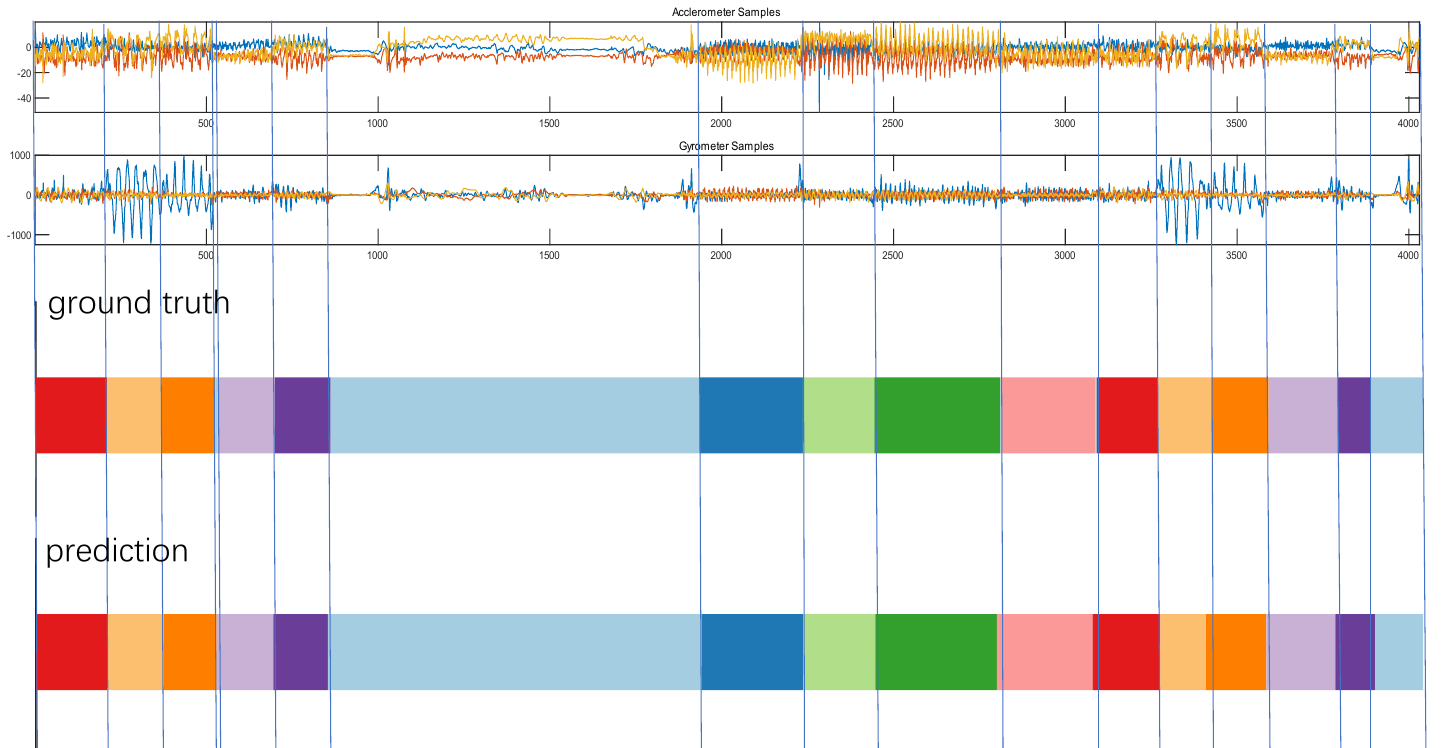}
    \caption{Results of the 1st subject in the laboratory hall.}
    \label{fig:hongli_1}
\end{figure*}

\begin{figure*}[t]
    \centering
    \includegraphics[width=1\linewidth]{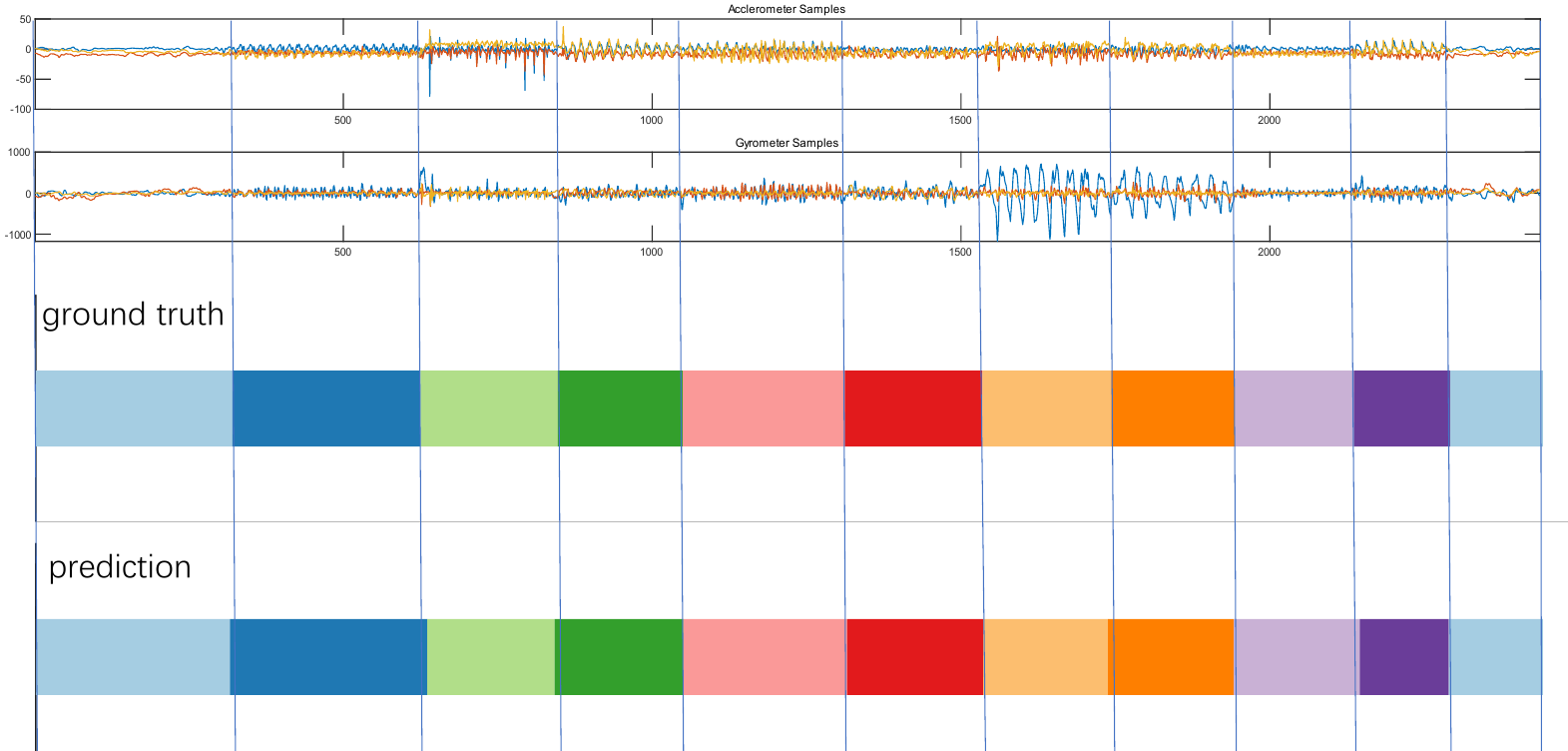}
    \caption{Results of the 7th subject in the laboratory hall.}
    \label{fig:hongli_7}
\end{figure*}

\begin{figure*}[t]
    \centering
    \includegraphics[width=1\linewidth]{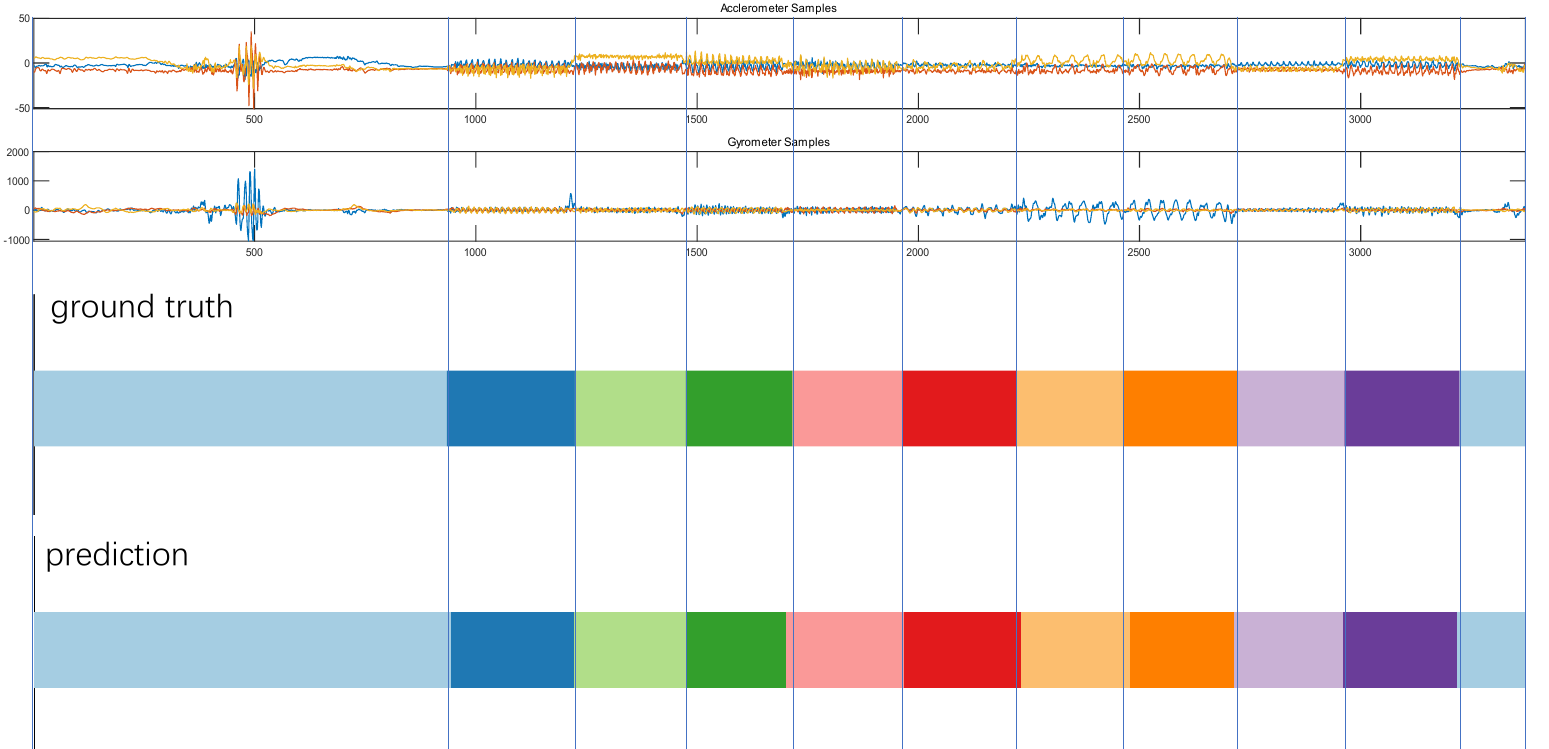}
    \caption{Results of the 1st subject in the library.}
    \label{fig:library_1}
\end{figure*}

\begin{figure*}[t]
    \centering
    \includegraphics[width=1\linewidth]{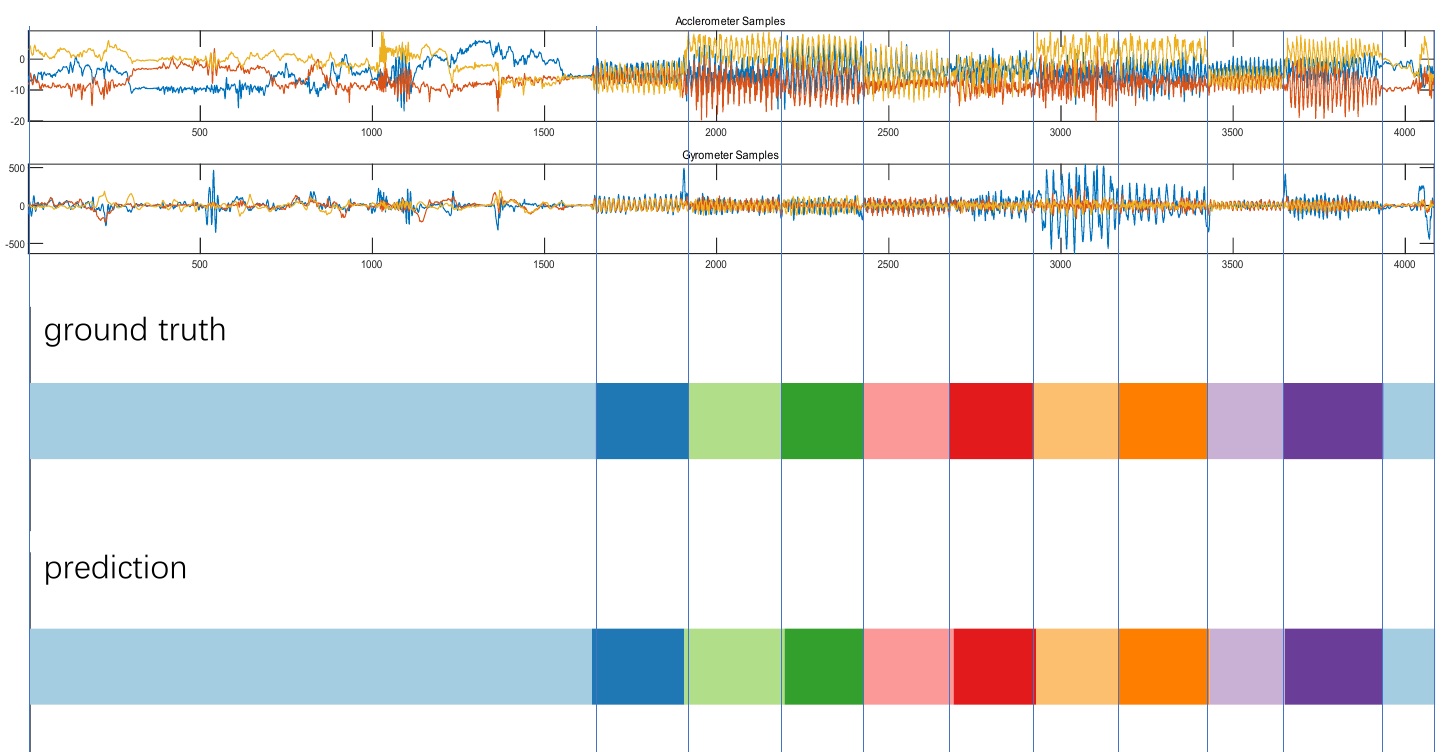}
    \caption{Results of the 2nd subject in the library.}
    \label{fig:library_2}
\end{figure*}

We show some results of UWash in Fig.~\ref{fig:canteen7}, Fig.~\ref{fig:canteen_10}, Fig.~\ref{fig:dormitory_4}, Fig.~\ref{fig:hanying_8}, Fig.~\ref{fig:hongli_1}, Fig.~\ref{fig:hongli_7}, Fig.~\ref{fig:library_1}, Fig.~\ref{fig:library_2}.

\end{document}